\documentclass[twocolumn,showpacs,preprintnumbers,amsmath,amssymb]{revtex4}

\usepackage{graphicx}
\usepackage{dcolumn}
\usepackage{bm}

\begin{document}

\preprint{APS/}

\title{Electron dynamics in the normal state of  cuprates: spectral function, Fermi surface  and ARPES data}

\author{E.E.  Zubov}
\affiliation{%
Institute for Metal Physics, National Academy of Sciences of Ukraine, Vernadsky Blvd. 36, Kyiv, 03680, Ukraine}%
\date{\today}

\begin{abstract}
An influence of  the electron-phonon interaction on excitation spectrum and damping in a narrow band electron subsystem of cuprates has been investigated. Within the framework of the t-J model  an  approach  to solving a problem of  account of  both  strong  electron correlations and local electron-phonon binding  with characteristic  Einstein mode $\omega _0$ in the normal state has been presented. In approximation Hubbard-I it was found an exact  solution to the polaron bands. We established that in the low-dimensional system with a pure kinematic part of Hamiltonian  a complicated  excitation spectrum is realized. It is determined mainly by peculiarities of the lattice Green's function.  In the definite area of the electron concentration and hopping integrals a correlation gap may be possible on the Fermi level. Also, in specific cases it is observed a doping evolution of  the  Fermi surface. We found that the strong electron-phonon binding enforces a degree of coherence of electron-polaron excitations near the Fermi level and spectrum along the nodal direction depends on wave vector module weakly. It corresponds to ARPES data. A possible origin of the experimentally observed kink in the nodal direction of cuprates is explained by fine structure of the polaron band to be formed near the mode -$\omega _0$.   
\end{abstract}

\maketitle

\section{\label{sec:level1}Introduction}

A description of the strongly correlated electron dynamics in cuprates is one of the intrigue task  in the theory of condensed matter physics. The history of investigations from the beginning of the discovery HTSC covers a considerable quantity of publications. Experimental tunnel, magnetic, resonance and ARPES data \cite{1,2,3} of pointed objects reflect the extremely complicated character of the interplay between charge, spin and lattice degrees of freedom. In spite of a wide experimental material as well as undoubted successful description of the cuprate electron structure by ab initio methods \cite{4,5} the series of observed phenomena to be connected with the strong electron correlations and electron-phonon interaction is not reproduced theoretically. In our opinion the reason is that within framework of such approaches  it is impossible to extract an effective self-consistent field by analogy with Weiss field in magnetism or unperturbed Hamiltonian for free electrons in metals as well as the restricted statistics and difficulties to account for hole states.   In some cases authors  use   the methods of  metal  physics especially to build the HTSC model and to account for strong electron-phonon interactions that is not useful for electron systems with strong  correlations \cite{6,7,8}. One of the puzzle phenomenon which is not described by existing theories is low-energy kink experimentally observed in the nodal dispersion of various cuprate materials \cite{9,10,11}. Below we will give a theory  which explains the origin of pointed anomaly  and presents the consecutive picture of the strongly correlated electron dynamics in cuprates.

In this connection it is necessary to point out that the aim of this work is to build  successive approximations for a diagrammatic method  when the Hubbard-I \cite{12} approximation is the starting point for any calculations. In this case we have a self-consistent field with the aid of which one can account for a specific character of doped Mott insulators. Indeed, below we will see the fundamental difference between metals and doped cuprates  resulting in a strong correlation narrowing of the valence band for hole doping \cite{13,14}. Within the framework of the t-J model for lower Hubbard band without exchange electron interaction it easy to detect that in paramagnetic phase (PM)  there are two noninteracting electron liquids with spin up and down avoiding one another and being in equilibrium. Hence, it gives rise  to correlation effect. Indeed, unlike metals here every electron hopping does not change a number of electrons for both sites with spin-up and spin-down, respectively. In metal every electron hopping gives site with double occupancy changing the fixed electron quantity in the pointed subsystems. That's why we must consider the whole electron system regardless of the spin.  As a result in PM phase of doped  Mott insulators in the limit of Coulomb repulsion $\textit{U} = \infty$ 
   the excitation spectrum will be two-fold degenerate and chemical potential   in the limit of half band filling is equal approximately to half of metal. It is necessary to point out that the ferromagnetically ordered  strongly correlated electron system is similar to ordinary metal because in this case the Fermi statistic is working first  of all.

Taking the strong electron-phonon interaction into account presents one of the most complicated task in the theory of condensed state. To solve this problem in a given work it is suggested to consider the Holstein model  with one characteristic optical Einstein mode. The extension  to case  of  few modes does not present any difficulties.  To solve problem we use the method of inverse function to be formulated by us in work \cite{15} for building the HTSC theory. Also, in contrast to existing theories \cite{16,17} where account of the  electron-phonon interaction can not be exact our model based on transformation of Lang-Firsov \cite{18} allows to solve this problem exactly. In this work in the Hubbard-I approximation first it has been shown  how a strong electron-phonon binding modifies the band spectrum, chemical potential and Fermi surface. To calculate the Green's  function we used  a such type of the Dyson's  equation when the self-energy depends on  frequency only and  all inhomogeneity is related exceptionally  to the  Fourier representation of  the hopping integral.

Based on the diagrammatic contributions related to inelastic electron scattering a new specific of the electron spectrum and spectral density is revealed when the frequency and wave characteristic are varied abruptly through a small range. It is in a good qualitative  agreement  with the ARPES data. This circumstance makes the Fermi surface detection difficult and in specific cases points out the non-Fermi liquid behaviour of the electron ensemble.

The structure of the paper is as follows. In section 2 the starting fermion-boson  Hamiltonian of the cuprate system is considered. It was diagonalized by unitary transformation. In approximation Hubbard-I  a detailed analysis of  the dynamic electron properties is carried out by taking into account  an influence of the electron-phonon interaction, next nearest neighbours and changing the Fermi surface topology. In section 3 the low-dimensional correlations and electron-phonon interactions  are included to provide the most general expression for  Matsubara Green functions in the first nonvanishing approximation of  time-dependent perturbation theory with respect to the inverse effective radius of interaction $\textit{r}$ $\approx$ $\textit{1/z}$, where $\textit{z}$ is the number of nearest neighbor in the simple square lattice. It was obtained the equations for excitation spectrum and damping. Also, the numerical analysis for chemical potential, Fermi surface, frequency spectrum and spectral density of  the PM phase at value of parameter of electron-phonon binding $\textit{g}$=0 was carried out. The polaron bands, spectral densities and their modifications versus parameter g were calculated. In conclusion we give the theoretical results having regard to the most typical experimental data.

\section{\label{sec:level1}Holstein polarons and effective self-consistent field in cuprates}

\subsection{\label{sec:level2} Hamiltonian of the fermion-bosonic system}

To describe the electron dynamics it is necessary to calculate the Matsubara electron Green's function  poles of which determine an excitation spectrum $\omega_{\textbf{\textit{k}}}$. Also, spectral density \textit{ A($\omega$,}\textbf{k}\textit{)} is proportional to the square power of  the imaginary part  of  Green's function. In what follows  we will consider a  PM phase in which the spin index $\sigma$ does not play any  role. The t-J model is believed to be the most simple for a description of the strong electron correlations. Apparently, in our case the exchange part of Hamiltonian is not considered. We add  to Hamiltonian of  t-J model the part used in the Holstein model  of  small polarons for the HTSC systems that are characterized by sufficiently strong electron-phonon interaction. These polarons are formed due to interaction between electrons and lattice optical  vibrations. For simplicity, it is considered the  Einstein model  with phonon frequency $\omega _0$. Thus, the Hamiltonian is written as 
\begin{eqnarray}
\hat{H}=\hat{H}_{0}+ V
\,,
\label{01}
\end{eqnarray}
where 
\begin{eqnarray}
\begin{array}{l} 
\hat{H}_{0} =\hat{H}_{0f} +\hat{H}_{b}  \\ {\hat{V}=-\sum\limits_{i,j,\sigma }t_{ij} c_{\sigma i}^{+} c_{\sigma j}  (1-n_{i-\sigma } )(1-n_{j-\sigma } )} \\ {\hat{H}_{0f} =-\mu \sum\limits_{i,\sigma }n_{i\sigma }  } \\ {\hat{H}_{b} =-g\sum\limits_{i}n_{i} \left(b_{i}^{+} +b_{i} \right)+\omega _{0} \sum\limits_{i}b_{i}^{+} b_{i}}    
\label{02}
\end{array}
\end{eqnarray} 

Here, the fermionic $\hat{H}_{0f} $ and bosonic $\hat{H}_{b} $ terms are  related to energy of chemical potential $\mu$ and local electron-phonon interaction in phonon subsystem with frequency $\omega _{0} $, respectively. The sum of these terms $\hat{H}_{0f} $ and $\hat{H}_{b} $  is considered as unperturbed Hamiltonian $\hat{H}_{0} $. For weakly doped cuprates  $\hat{V}$ operator is taken as the perturbation, where both $c_{\sigma i}^{+} \; (c_{\sigma i}^{} )$ and $b_{i}^{+} \; (b_{i}^{} )$ create (annihilates) an electron of spin $\sigma$ and phonon on lattice site $\textit{i, }$ respectively, and  $\textit{t} _{ij}$ is the hopping integral. $n_{i} =n_{i\sigma } +n_{i-\sigma } $is the site electron concentration, where $n_{i\sigma }$ represents the concentration of electrons on site \textit{i  }with spin $\sigma$. The exchange term of the Hamiltonian is neglected for the system in PM state. 

The Lang-Firsov unitary transform \cite{18} allows to separate the boson and fermion operators in $\hat{H}_{b} $. Then the transformed $\hat{H}_{b} $ takes the form $\hat{\tilde{H}}_{b} $:

\begin{eqnarray}
\hat{\tilde{H}}_{b} =\omega _{0} \sum\limits_{i}b_{i}^{+} b_{i}  -\xi \sum\limits_{i}n_{i}
\,,
\label{03}
\end{eqnarray}
where $\xi =g^{2} /\omega _{0} $ is the polaron binding energy. Also, perturbation Hamiltonian $\hat{V}$ is transformed to $\hat{\tilde{V}}$:

\begin{eqnarray}
\hat{\tilde{V}}=-\sum _{<ij>,\sigma }t_{ij} \tilde{c}_{i\sigma }^{+} \tilde{c}_{j\sigma }  (1-n_{i-\sigma } )(1-n_{j-\sigma } )
\,,
\label{04}
\end{eqnarray}
Here, the unitary transformated   Fermi operators
\begin{eqnarray}
\tilde{c}_{i\sigma } =Y_{i} c_{i\sigma }
\,,
\label{05}
\end{eqnarray}
are product of Bose   $Y_{i} =e^{\lambda (b_{i} ^{+} -b_{i} )} \quad $ and corresponding Fermi destruction operators  where \textit{$\lambda$=g/$\omega_0$}.  Another  terms in Hamiltonian \eqref{01} are not transformed.

Therefore we have unperturbed Hamiltonian with perturbation  $\hat{V}$ that allows to build an approach  based on the scattering matrix formalism.

\subsection{\label{sec:level2} Approximation Hubbard-I}

In work \cite{12} J. Hubbard  developed a simplest approach to decoupling of the correlators  in equation of motion for Green's function. It describes the dielectric state of the strongly correlated electron system. Although this approach denoted as approximation Hubbard-I do not describe a correlation-induced metal-insulator transition it is quite similar to Weiss mean field theory of magnetism with  an effective field and therefore has fundamental importance in understanding of  the  physics of  strong electron correlations. And so in what follows we will consider  this approach in detail within the framework of modern diagrammatic  method  taking into account that a such systematic description is absent in literature. It is necessary  to point out that this method is a quite equivalent to approach to be used by J. Hubbard.

 Below we will consider the lower Hubbard's band doped by holes. In this case one can use one hole and two electron site wave functions: ${\left| \psi _{0}  \right\rangle} ={\left| 0 \right\rangle} ,\, {\left| \psi _{\sigma }  \right\rangle} ={\left| \sigma  \right\rangle} $, where $\sigma$=+ (or 1)  and   $\sigma$=- (or -1)  for electrons with spin-up and -down, respectively. Let us introduce the Hubbard's operators $X^{ik} ={\left| \psi _{i}  \right\rangle} {\left\langle \psi _{k}  \right|} $ in a given basis. Apparently, in the projective space for the destruction and creation  operators $c_{\sigma j} (1-n_{j-\sigma } )=X_{j}^{^{0\sigma } } $ and $c_{\sigma j}^{+} (1-n_{j-\sigma } )=X_{j}^{^{\sigma 0} } $, respectively. With account of  Eqs.\eqref{03}-\eqref{05} the Hamiltonian of the subsystem is written as
\begin{eqnarray}
\hat{H}_{el.} =\hat{\tilde{H}}_{0} +\tilde{V}
\label{06}
\end{eqnarray}
where the unperturbed Hamiltonian takes the form:
\[\hat{\tilde{H}}_{0} =-\tilde{\mu }\sum _{i}n_{i}  ,                                                                                  \] 
and  perturbation appears as

\begin{eqnarray}
\tilde{V}=-\sum _{<ij>,\sigma }t_{ij} \tilde{X}_{i}^{\sigma 0} \tilde{X}_{j}^{0\sigma }
\label{07}
\end{eqnarray}
where     $\tilde{\mu }=\mu +\xi $    is effective chemical potential with account of electron-phonon interaction,   $\tilde{X}_{i}^{\sigma 0} =Y_{i}^{+} X_{i}^{\sigma 0} $ and   $\tilde{X}_{j}^{0\sigma } =Y_{j} X_{j}^{0\sigma } $  are the unitary transformed  Hubbard's operators. In spite of  the simple form the Hamiltonian \eqref{06} contains practically all complicated  electron dynamics with lattice coupling. It will be seen below from the given calculations.

In the context of the  time-dependent perturbation theory it is easy to show that equation for the Fourier transform $\Lambda _{0\sigma } (i\omega _{p} ,{\textbf q})$ of the Matsubara Green's function $\Lambda _{0\sigma } ({\rm l}\tau _{{\rm l}} ,{\rm m}\tau _{{\rm m}} ){\rm =}{\rm -<T}_{\tau } \tilde{X}_{l}^{0\sigma } (\tau _{l} )\tilde{X}_{m}^{\sigma 0} (\tau _{m} )>$ in approximation Hubbard-I  is presented graphically in form as in Fig.1, where the symbol  ${\rm <}...>$ denotes a statistical averaging over the total Hamiltonian $\hat{H}_{el.} $ $\tilde{X}_{l}^{\alpha \beta } (\tau _{l} )$ and ${\rm T}_{\tau } $ are Hubbard's operators in interaction representation  and time-odering  operator, respectively. Here, the bold and  thin  lines correspond to $\beta {\kern 1pt} \Lambda _{0\sigma } (i\omega _{p} ,{\textbf q})$ and $\beta {\kern 1pt} \tilde{G}_{0\sigma } (i\omega _{p} )<F^{\sigma 0} >$, respectively, where \textit{1/$\beta$=T} is temperature, $\omega _{p} =(2p+1)\pi /\beta $, $<F^{\sigma 0} >=1-n/2$ is the mean electron-hole probability of  site occupancy for electron concentration \textit{n}. The unperturbed Green's function \cite{14,15}
\begin{eqnarray}
\tilde{G}_{0\sigma } (i\omega _{p} )= \frac{e^{-\lambda ^{2} } }{\beta }   \kern 70pt    
\nonumber \\ \times\sum _{m=0}^{\infty }\frac{\lambda ^{2m} }{m!} \left\{\frac{f(-\tilde{\mu })}{i\omega _{p} +\tilde{\mu }+m\omega _{0} } +\frac{1-f(-\tilde{\mu })}{i\omega _{p} +\tilde{\mu }-m\omega _{0} } \right\}
\, ,
\label{08}
\end{eqnarray}
Here, $f(x)=1/(e^{\beta x} +1\; )$ is the Fermi distribution. We suppose that \textit{$\omega_0$/T}$>$$>$1. A wave line in Fig.1 denotes the Fourier transform 
\[\begin{array}{c}
t(\textbf{q})=\sum\limits_{i,j}t_{ij} e^{-i\textbf{q}(\textbf{r}_{i} -\textbf{r}_{j} )} =  \\  -2t(\cos (q_{x} a)+\cos (q_{y} a)+2\alpha \cos (q_{x} a)\cos (q_{y} a)) =2t\varepsilon
\end{array}
\]
of the hopping integral, where \textit{$\alpha $}= \textit{$t_1/t$},  \textit{t}  and \textit{$t_1$} are nearest and next nearest hopping integrals,  respectively, for square lattice with parameter \textit{a}. We have electron excitations  if  the parameter \textit{t$>$0} and  hole spectrum otherwise. The replacement \textit {t $\to$ -t} and \textit{ $\alpha \to$} -\textit{$\alpha $}  has no influence on a position of  the Fermi level.

 In approximation Hubbard-I  a mean $<F^{\sigma 0} >$ is the self-consistent parameter. Hence, we make the replacement $<F^{\sigma 0} >_{0} \to <F^{\sigma 0} >$, where 
\begin{eqnarray}
<F^{\sigma 0} >_{0} =<X^{\sigma \sigma } +X^{00} >_{0}\kern 70pt    
\nonumber \\ =\frac{Sp(\exp (-\beta \hat \tilde{H}_{0} (X^{\sigma \sigma } +X^{00} )))}{Sp(\exp (-\beta \hat \tilde{H}_{0} ))} =\frac{e^{\beta \tilde{\mu }} +1}{1+2e^{\beta \tilde{\mu }} }
\label{09}
\end{eqnarray}
\begin{figure}[h]
\begin{center}
\includegraphics[width=\columnwidth]{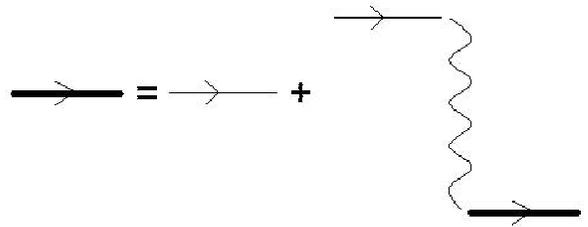}
\end{center}
\caption{The  graphic equation for Green's function $\Lambda _{0\sigma } (i\omega _{p} ,{\textit{\textbf q}})$ in approximation Hubbard-I.}
\label{fig.1}
\end{figure}
The solution of graphic equation in Fig.1 is

\begin{eqnarray}
\beta {\kern 1pt} \Lambda _{0\sigma } (i\omega _{p} ,\textbf q)=\frac{\beta {\kern 1pt} \tilde{G}_{0\sigma } (i\omega _{p} )<F^{\sigma 0} >}{1-\beta t(\textbf q)\tilde{G}_{0\sigma } (i\omega _{p} )<F^{\sigma 0} >}
\label{10}
\end{eqnarray}
The retarded Green's functions  are obtained  by the analytic continuation $i\omega _{n} \to \Omega +i\delta $ in \eqref{10} that gives  a full information  about electron dynamics within the framework of the approximation Hubbard-I. For simplicity we assume that \textit{T} = 0. By introducing the symbol $w=(\Omega +\tilde{\mu })/\omega _{0} $  we have for unperturbed Green's function \eqref{08}
\begin{eqnarray}
\beta \tilde{G}_{0\sigma } (w\omega _{0}-\tilde{\mu})=\frac{1}{{w\omega _0}} ( M(1,1+w,-\lambda ^{2} )\theta (\tilde{\mu}) \kern 30pt    
\nonumber \\   
+M(1,1-w,-\lambda ^{2} )\theta (-\tilde{\mu })), \kern 20 pt
\label{11}
\end{eqnarray}
where  \textit{M(a,b,z)} and $\theta (x)$ are\textit{ } the confluent hypergeometric function of Kummer \cite{20} and  Heaviside step function, respectively.

It is easy to find the pole singularities based on  Eq.\eqref{10} that gives the spectrum of  electron-hole excitation in approximation Hubbard-I:
\begin{eqnarray}
\begin{array}l
\varepsilon _{c.m}  =  4\left( {\frac{{t(\textbf{k})}}{W}} \right)_{c.m}  =  \frac{{4(\Omega _{\textbf{k}m}  + \tilde \mu )}}{{ < F^{\sigma 0}  > M\left( {1,1 + \frac{{\Omega _{\textbf{k}m}  + \tilde \mu }}{{\omega _0 }}, - \lambda ^2 } \right)}}  \\  = -(\cos (k_x a) + \cos (k_x a) + 2\alpha \cos (k_x a)\cos (k_x a)),
\label{12}
\end{array}
\end{eqnarray}
where bandwidth \textit{W=8t$>$0}. Index  \textit{m}  enumerates the modes which are the solutions of Eq.\eqref{12}. Let  \textit{W}=1, i.e. the frequencies, chemical potential and all energy parameters are measured in the units of  bandwidth.

In Fig.2 the \textit{k} dependence of  resonance  frequency $\Omega _{{\it k}} $ is presented. The curves were calculated  along nodal direction \textit{$k_x$}= \textit{$k_y$}=$k/\sqrt{2}$ at a$=$3.814 $\mathop {\rm A}\limits^o$ at different electron concentrations and electron-phonon coupling   for bismuth cuprate  2212 with phonon frequency \textit{$\omega _0$}= 0.01875. From figure it easy to see that the inclusion of the electron-phonon coupling  (\textit{g}$\neq$0) results in the appearance of  polaron bands the bandwidth of which is increasing with increasing \textit{g}.  It follows from   Eq. \eqref{12}  that  kinks on curve  $\Omega _{{\it k}} $ versus \textit{k} are determined by zeros of function $M\left(1,1+\frac{\Omega _{{\it k}{\it m}} +\tilde{\mu }}{\omega _{0} } ,-\lambda ^{2} \right)$. Computing the real zeros of hypergeometric functions is a complicated mathematical  problem which can be solved only numerically \cite{21}. However, one can state that the centre of the envelope curve of   polaron  bands  and their total bandwidth  are determined  approximately by parameter $\lambda ^{2} $ which counts the number of phonon quanta in the phonon cloud around the localized electron \cite{1}, i.e. at $\frac{\Omega _{{\it k}{\it m}} +\tilde{\mu }}{\omega _{0} } \sim -\lambda ^{2} $. For high electron energies the polaron band is localized and its energy is equal to energy of  phonon quanta in this area of frequencies.
\begin{figure}[h]
\begin{center}
\includegraphics[width=\columnwidth]{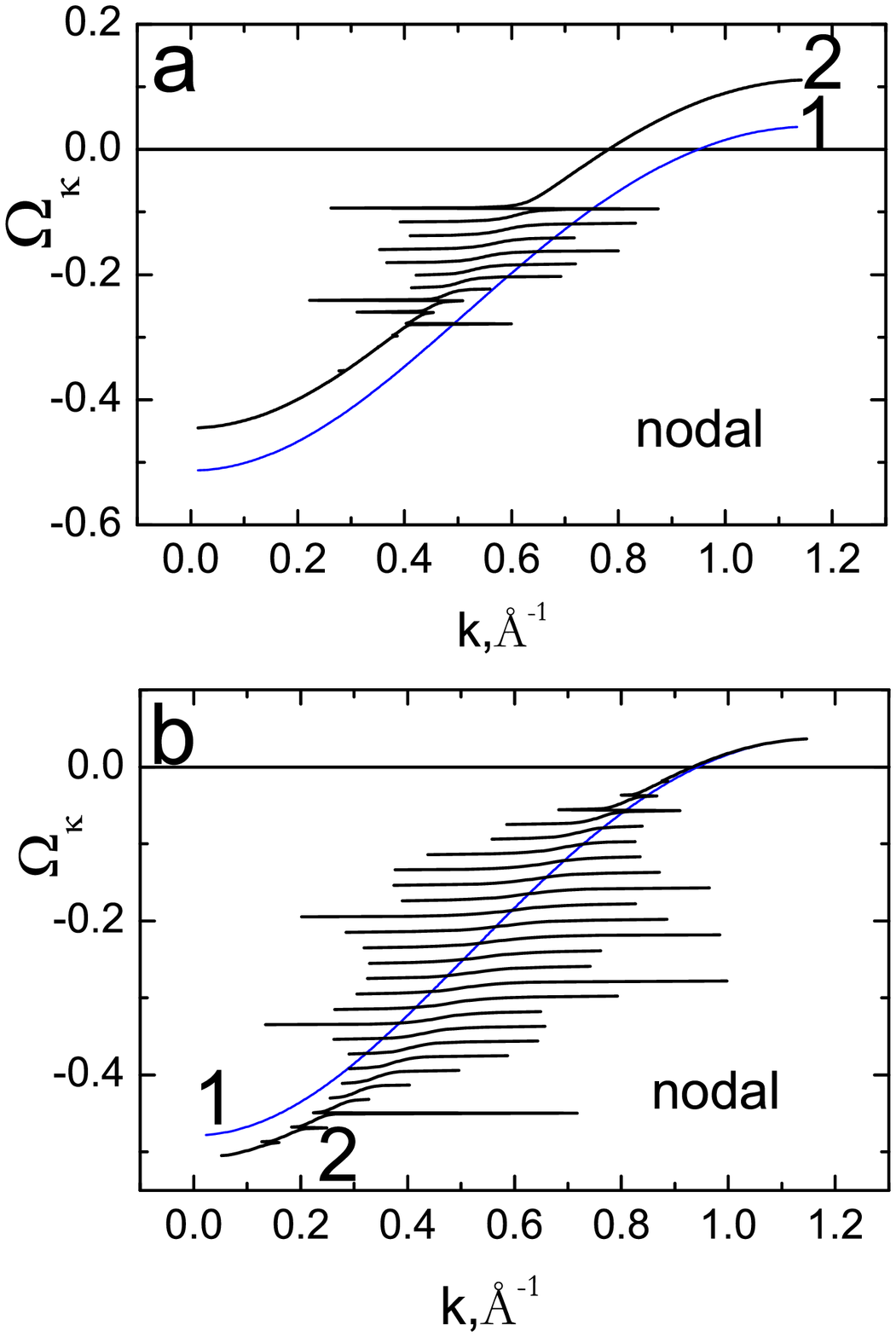}
\end{center}
\caption{The electron excitation frequency-momentum dispersion along the nodal direction  \textit{$k_x = k_y = k/$}$\sqrt{2}$  at  a$=$3.814 $\mathop {\rm A}\limits^o$ for bismuth cuprate 2212 with phonon frequency   $\omega_0$= 0.01875  and  a)  \textit{n}=0.9, $\alpha $=0.1, \textit{g}=0, $\tilde{\mu }$=0.136 and \textit{g}=0.03, $\tilde{\mu }$=0.092 (curves 1 and 2, respectively); b)  \textit{n}=0.97, $\alpha $=0.1, \textit{g}=0, $\tilde{\mu }$= 0.195 and \textit{g}=0.06, $\tilde{\mu }$=0.0188 (curves 1 and 2, respectively). The bandwidth \textit{W}=1 hereinafter.}
\label{fig.2}
\end{figure}

Let us consider the case when polarons are absent, i.e. at $\lambda =0$. Then $M\left(1,1+\frac{\Omega _{{\it k}{\it m}} +\tilde{\mu }}{\omega _{0} } ,-\lambda ^{2} \right)$=1 and electron excitation spectrum consists from one mode and has the form:
\begin{equation} 
\label{13} 
\begin{array}{l}
 \Omega _k  + \tilde \mu  = \frac{{ - 1}}{4} < F^{\sigma 0}  >  \\ 
  \times \left( {\cos (k_x a) + \cos (k_x a) + 2\alpha \cos (k_x a)\cos (k_x a)} \right) \\ 
 \end{array}\end{equation} 

The spectrum \eqref{13}  has a characteristic feature which radically differentiates the metal state from state of doped  Mott's dielectrics. Indeed, the site probability ${<F^{\sigma 0} >}$ of the electron-hole state enters in Eq.\eqref{13}. In general case of ordered spins $<{F^{\sigma 0} >}={1-n/2+\sigma <S^{z}}>$. In ferromagnetic phase  a  mean spin $<S^{z}>=n/2$ that gives $<F^{+0}>=1$ and $<F^{-0}>=1-n$. Thus, in the limit of half filled band  at $n\to 1$ we have practically one band in  accordance with Eq.\eqref{13}. It is typical  for metal state. On the other hand, in PM state $<S^{z} >=0$, $<F^{+0}>=<F^{-0}>=1-n/2$, i.e. mode \eqref{13} is two-fold degenerate. This is because in this case spin-up and spin-down electron liquids  coexist in cuprates. These liquids  do not  interact because their Hamiltonians \eqref{07} commute. Apparently, the Fermi level must be shifted  essentially since  the electron liquids are in thermodynamic equilibrium. This factor should be taken into account especially when the methods of metal theory  are applied to hole doped Hubbard's  systems. Let us dwell on this problem in detail by the example of  calculation of chemical potential value. 

To solve problem it is necessary to find the electron density of state which is determined by imaginary  part of the lattice Green's function analytically continuated  to the lower  half-plane \cite{22}:
\begin{eqnarray}
 G(s,\alpha ) \kern 100pt   \nonumber 
\\
  = \frac{1}{N}\sum\limits_\textbf{k} {\frac{1}{{s - i\delta  - cos(k_x a) - (1 + 2\alpha cos(k_x a))cos(k_y a)}}} \nonumber \\ 
\,
\label{14} 
\end{eqnarray}
The given sum is calculated exactly. Taking into account that 2$\alpha $$<$1 one can write the real and imaginary parts of $G(s,\alpha )$ for \textit{s}$>$0:

\begin{widetext}
\begin{eqnarray}
\label{15} 
ReG(s,\alpha )=\left\{\begin{array}{l} {\frac{2}{\left|s+2\alpha \right|} \frac{1}{\pi } K\left(\frac{2\sqrt{2\alpha s+1} }{\left|s+2\alpha \right|} \right),\quad 2\alpha s+1>0,\, s>0,\; s>2(\alpha +1)} \\ {\frac{1}{\sqrt{2\alpha s+1} } \frac{1}{\pi } K\left(\frac{\left|s+2\alpha \right|}{2\sqrt{2\alpha s+1} } \right),\quad 2\alpha s+1>0,s>0,\, \, 2(\alpha -1)\; \le s\le 2(\alpha +1)} \\ {\frac{2}{\sqrt{(2\alpha -s)^{2} -4} } \frac{1}{\pi } K\left(2\sqrt{\frac{-2\alpha s-1}{(2\alpha -s)^{2} -4} } \right),\quad 2\alpha s+1\le 0,\, s>0,\; s>2(\alpha +1)} \end{array}\right.\end{eqnarray}
\begin{eqnarray}
\label{16}
ImG(s,\alpha )=\left\{\begin{array}{l} {\frac{1}{\sqrt{2\alpha s+1} } \frac{1}{\pi } K\left(\frac{1}{2} \sqrt{\frac{4-(2\alpha -s)^{2} }{2\alpha s+1} } \right),\quad \; 2(\alpha -1)\; <s<2(\alpha +1)} \\ {0,\quad otherwise} \end{array}\right.  ,       
\end{eqnarray}
\end{widetext}
where \textit{K(k)} is a complete elliptic integral of the first order. We point out that \textit{G(s,$\alpha $)} has a such symmetric property:
\begin{eqnarray}
G(-s+i\delta ,\alpha )=-G(s-i\delta ,-\alpha )
\label{17} 
\end{eqnarray}
Using Eq.\eqref{17} we write the next symmetric property for real part of  lattice Green's function:
\begin{eqnarray}
ReG(s,-\alpha )=-ReG(-s,\alpha )
\label{18} 
\end{eqnarray}
The pointed property allows to consider an electron spectrum type of Eq.\eqref{13} since a determination of the  lattice Green's function corresponds to hole spectrum for  $s>0$. In what follows for calculations we use the left part of Eq.\eqref{18} and keep in mind that values -$\alpha$ and $\alpha$  correspond to electrons and holes, respectively. Hence, in expressions for lattice Green's functions in the electron spectrum with parameter $\alpha$ we will take $ReG(s,-\alpha )$. 

The imaginary part required for further calculations is only determined by analytic continuation  in a lower half-plane independently of  replacement $s\to -s$. As a result we have the next  relationship for imaginary part of $G(s,\alpha )$:
\begin{eqnarray}
ImG(-s,\alpha )=ImG(s,-\alpha )
\label{19} 
\end{eqnarray}
Eqs.\eqref{18} and \eqref{19} allow to find the real and imaginary parts of $G(s,\alpha )$ for all values of  \textit{s}.

In Fig.3 the dependences on parameter  \textit{s} for  real (a) and imaginary (b) parts of the lattice Green's function \eqref{14}  at  \textit{$\alpha $}=0, -0.4 and 0.4 (curves 1-3, respectively) are presented. From figure one can see the asymmetry of curves and van Hove singularities  for nonzero values of parameter \textit{$\alpha $} that is typically for two-dimensional systems. For \textit{s}=0 $ReG(s,\alpha )$ has  discontinuity. It has to give rise to qualitative changing in the spectrum near the point  \textit{s}=0.
\begin{figure}[h]
\begin{center}
\includegraphics[width=\columnwidth]{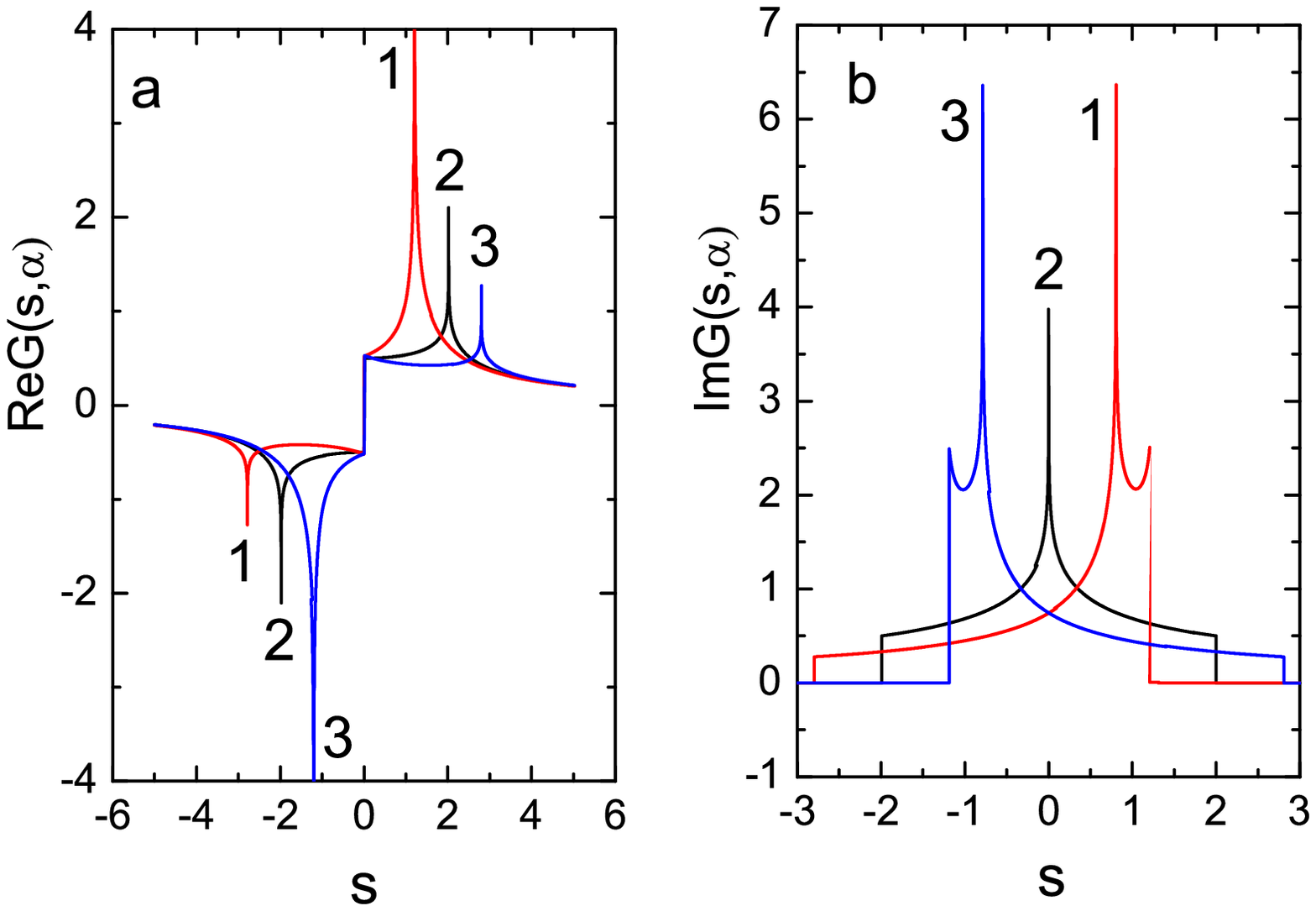}
\end{center}
\caption{The \textit{s}-dependences  of  real (a) and imaginary (b) parts of the lattice Green's function \eqref{14}  at  \textit{$\alpha $}=0, -0.4 and 0.4 (curves 1-3, respectively).}
\label{fig.3}
\end{figure}

From Eq.\eqref{15} it follows that for transition from hole to electron consideration the sign of  a resonance frequency is changed on opposite. The equation for chemical potential is determined by $ImG(s,\alpha )$ which does not depend on hole or electron formalism.
One can make the summation over all wave vectors for any function \textit{V($\varepsilon $)} by the use of a  lattice Green's function. Indeed, we have
\begin{eqnarray*}
 S(\alpha ) = \frac{1}{N}\sum\limits_\textbf{q} {V(cos(k_x a) + (1 + 2\alpha cos(k_x a))cos(k_y a))}  \\ 
  = \frac{1}{N}\sum\limits_\textbf{q} {\int\limits_{ - 2 + 2\alpha }^{2 + 2\alpha } {\delta ( - \varepsilon  - cos(k_x a) - (1 + 2\alpha cos(k_x a))} }  \\ 
  \times cos(k_y a))V( - \varepsilon )d\varepsilon  = \int\limits_{ - 2 - 2\alpha }^{2 - 2\alpha } {D_C (\varepsilon ,\alpha )V(\varepsilon )d\varepsilon },
 \end{eqnarray*}
where \textit{$\delta $(x)} is the Dirac delta function. The electron density of state $D_{C} (\varepsilon ,\alpha )$ has the form:
\begin{eqnarray}
 D_C (\varepsilon ,\alpha )\nonumber \kern 200pt \\
  = \frac{1}{N}\sum\limits_\textbf{q} {\delta (\varepsilon  - cos(q_x a) - (1 + 2\alpha cos(q_x a))cos(q_y a))} \nonumber  \\ 
  = \frac{1}{\pi }{\mathop{\rm Im}\nolimits} G(\varepsilon ,\alpha )\kern 100pt
\label{20}
\end{eqnarray}
Apparently, $D_{C} (\varepsilon ,\alpha )$ has nonzero values for $\varepsilon $ in area from -2-2$\alpha $  to  2-2$\alpha $.

From Eq.\eqref{10} for Green's function $\beta {\kern 1pt} \Lambda _{0\sigma } (i\omega _{n} ,\textbf{q})$ in approximation Hubbard-I we find the spectral density of  the fermion-bosonic system:
\begin{eqnarray}
A_{Hubb.} (\Omega ,{\textbf{k}})=-2\beta Im\Lambda _{0\sigma } (\Omega +i\delta ,{\textbf{k}})
\label{energy21} 
\end{eqnarray}
Expanding the denominator of $\Lambda _{0\sigma } (\Omega +i\delta ,{\textbf{k}})$ in a series in the vicinity of its \textit{m}-th pole $E_{n{\textbf{k}}\sigma } $  we have
\begin{eqnarray}
\label{22} 
A_{Hubb.} (\Omega ,{\it \textbf{k}})=-2\pi \sum _{m}\frac{\beta \tilde{G}_{0\sigma } (E_{{\it m\textbf{k}}\sigma } )}{t({\textbf{k}})\frac{d\beta \tilde{G}_{0\sigma } (\Omega )}{d\Omega } } \delta (\Omega -E_{{\it m\textbf{k}}\sigma } ) 
\end{eqnarray}
Here, the sum is over  \textit{m}-th modes $E_{m{\textbf{k}}\sigma } $that are the implicit solutions of dispersion Eq. \eqref{12}. It is evident from Eq.\eqref{22} that  approximation Hubbard-I describes coherent excitation only. Having calculated   spectral density it is  easy to find  the equation for chemical potential. Indeed, a mean site occupancy  $<X^{\sigma {\kern 1pt} \sigma } >$ of electron with spin  $\sigma $  is determined by $A_{Hubb.} (\Omega ,{\textbf{k}})$:
\begin{eqnarray}
<X^{\sigma {\kern 1pt} \sigma } >=\frac{1}{N} \sum _{{\textbf{q}}}\frac{1}{2\pi } \int _{-\infty }^{+\infty }f(\Omega )A_{Hubb.} (\Omega ,{\textbf{q}})  d\Omega 
\label{23} 
\end{eqnarray}

In work \cite{15}  the method of inverse function was suggested to compute the integrals like that in Eq.\eqref{23} when an infinite number of modes $E_{n{\textbf{k}}\sigma } $ cannot be expressed as explicit. For that let us introduce the notation $\beta \tilde{G}_{0\sigma } (\Omega )=F(\Omega )$. The complicated delta function from Eq.\eqref{22} is simplified by relationship \cite{19}:
\begin{eqnarray}
W\delta [\Omega -E_{{\it m}{\textbf{q}}\sigma } ]=\frac{\delta (\varepsilon -\varepsilon _{c.m} )}{\left|(\Omega -E_{{\it m\textbf{q}}\sigma } )^{'} \right|_{\varepsilon _{c.m} } },
\label{24} 
\end{eqnarray}
where $\varepsilon _{c.m} =\frac{4}{F(\Omega )<F^{\sigma 0} >} $ and index \textit{m }gives  the range  $m{\kern 1pt} {\kern 1pt} \omega _{0} <\Omega <(m+1){\kern 1pt} {\kern 1pt} \omega _{0} $  in which the inverse function $F^{ - 1} \left( \varepsilon  \right)$ is determined. Apparently, that $E_{nq\sigma } (\varepsilon ) = F^{ - 1} \left( {\frac{4}{{\varepsilon  < F^{\sigma 0}  > }}} \right)$ and $E_{nq\sigma } (\varepsilon _{c.m} )=\Omega $. Taking into account a differentiation of  the inverse function   $F^{ - 1} \left( \varepsilon  \right)$ we have
\begin{eqnarray}
\left|(\tilde{\Omega }-\tilde{E}_{{\it m}{\textbf{q}}\sigma } (\varepsilon ))^{'} \right|_{\varepsilon =\varepsilon _{c.{\it m}} } =\left. \frac{4}{\varepsilon ^{2} <F^{\sigma 0} >\left|F^{'} (\varepsilon )\right|} \right|_{\varepsilon =\varepsilon _{c.{\it m}} }
\label{25} 
\end{eqnarray}
Substituting  \eqref{25} in \eqref{24}  and \eqref{24} in \eqref{23} with account for $\left|F'(\Omega )\right|/F'(\Omega )=-1$ we obtain
\[
\begin{array}{l}
  < X^{\sigma {\kern 1pt} \sigma }  >  = \sum\limits_{m = 0}^\infty  {\int\limits_{m{\kern 1pt} \omega _0 }^{(m + 1)\omega _0 } {d\Omega f(\Omega )\int\limits_{ - 2 - 2\alpha }^{2 - 2\alpha } {d\varepsilon {\kern 1pt} \varepsilon D_c (\varepsilon ,\alpha )} } }  \\ 
  \times \;\delta (\varepsilon  - \varepsilon _{c.m} )F(\Omega ) < F^{\sigma 0}  >,
 \end{array}
\]
that gives the next expression for mean site occupancy with electron-phonon binding:
\begin{eqnarray}
<X^{\sigma {\kern 1pt} \sigma } >=4\int _{-\infty }^{+\infty }d\Omega {\kern 1pt} f(\Omega )D_{c} \left(\frac{4}{F(\Omega )<F^{\sigma 0} >} ,\alpha \right)
\label{26} 
\end{eqnarray}

At temperature \textit{T}=0 in PM phase $<X^{\sigma {\kern 1pt} \sigma } >=n/2$ and from Eq.\eqref{26} we obtain the equation for chemical potential $\tilde{\mu }$:
\begin{eqnarray}
\frac{n}{2} = 4{\kern 1pt} {\kern 1pt} \omega _0 \nonumber \kern 150pt \\ 
  \times \int\limits_{ - \infty }^{\tilde \mu /\omega _0 } {dw{\kern 1pt} D_c \left( {\frac{{4w{\kern 1pt} {\kern 1pt} \omega _0 }}{{M(1,1 + w, - \lambda ^2 )\left( {1 - 0.5n} \right)}},\alpha } \right)}
\label{27}
\end{eqnarray}

From \eqref{27} it follows that in the absence of electron-phonon interaction the chemical potential is expressed  in form \cite{15}:
\begin{eqnarray}
\tilde{\mu }=\frac{2-n}{8} I^{-1} \left(\frac{n}{2-n} ,\alpha \right),
\label{28}
\end{eqnarray}
where  $I^{-1} \left(x,\alpha \right)$ is the inverse function of function $I(x,\alpha )=\int _{-2-2\alpha }^{x}D_{C} (\varepsilon ,\alpha )d\varepsilon  $. In particular, we have $I^{-1} \left(1,\alpha \right)=2-2{\kern 1pt} {\kern 1pt} \alpha $ and $\tilde{\mu }=\frac{1}{4} (1-\alpha )$ for \textit{n}=1. Thus, in PM phase the level of  chemical potential  tends to  the middle of upper half of  a nearly half-filled  band but not to band edge as it is required for normal metal. This correlation effect narrows the lower Hubbard's band by reason of  existence of  two spin-up and spin-down electron  liquids which are avoiding one another. For the first time this phenomena was observed by  W.F. Brinkman and  T.M.Rice \cite{13}.

In Fig.4 the concentration dependencies of the effective chemical potential $\tilde{\mu }$ evaluated from Eq.\eqref{27} at \textit{g}=0 for different values of  $\alpha $ (a), at $\alpha $=0 for different values of  \textit{g} (b) and the phase diagram in coordinate \textit{g-n} for $\alpha $=0 and  0.1 (curves 1 and 2,respectively) are presented. From Fig.4 a it easy to see that  for  \textit{n}$>$0.87 the chemical potential is increased with  decrease the value of \textit{$\alpha $}. For \textit{n}$<$0.87 in area of positive $\tilde{\mu }$  the inverse  trend is observed. From Fig.4  b one sees that with increasing \textit{g} the area of existence PM phase narrows and for \textit{g}=0.072 a dielectric state is realized (see Fig.4 c). Thus, the effective chemical potential $\tilde{\mu }$  is decreased up to zero with increasing \textit{g}, i.e. Fermi level is shifted  to band  centre.
\begin{figure}[h]
\begin{center}
\includegraphics[width=\columnwidth]{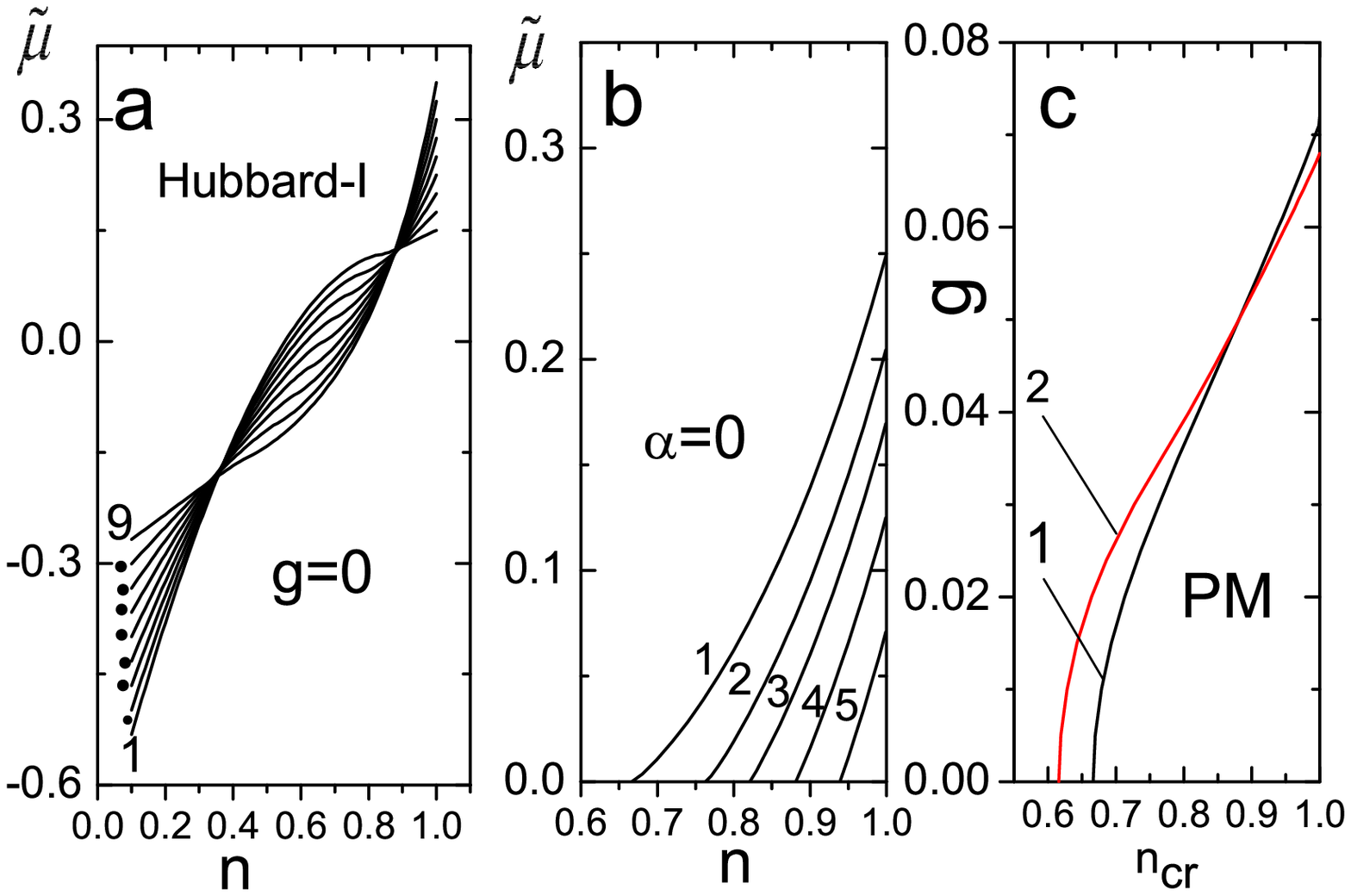}
\end{center}
\caption{The concentration dependences of the effective chemical potential $\tilde{\mu }$ in approximation Hubbard-I  without and with  account for polaron excitations for $\omega_0$= 0.01875 at  a) g=0 and $\alpha $=0.4, 0.3, 0.2, 0.1,0, -0.1, -0.2, -0.3 è -0.4 (curves 1-9, respectively);  b) $\alpha $=0 and g=0, 0.03, 0.04, 0.05 è 0.06 (curves 1-5, respectively); ñ) phase diagrams for $\alpha $=0 and 0.1 (curves 1 and 2, respectively).}
\label{fig.4}
\end{figure}

In Fig.5 the critical values of -\textit{$\epsilon_{cr.}$} from Eq.\eqref{12}  versus \textit{n} on Fermi level when $\Omega _{{\it k}{\it m}} =0$ at \textit{g}=0.03, 0.05 and 0.06  are shown. One sees that near the PM phase boundary when $\tilde{\mu }\sim 0$  the antibonding orbitals of second- nearest neighbors become important. That's why in this area  of electron concentration   the Fermi surface has a hole origin with the centre in  $\Gamma$ point (see Fig.6). As \textit{n} is increased  the electron  Fermi surface is arisen. Fig.6 reflects the topology dependence of the electron Fermi surface on both electron-phonon binding and doping. One shrink as doping is decreased. With increase the constant of an electron-phonon interaction g this shrink is decreased. Also, when the hole doping goes to zero the Fermi surface does not vanish.  Hence, the transition to dielectric state implies discontinuous   disappearance of the Fermi surface.
\begin{figure}[h]
\begin{center}
\includegraphics[width=\columnwidth]{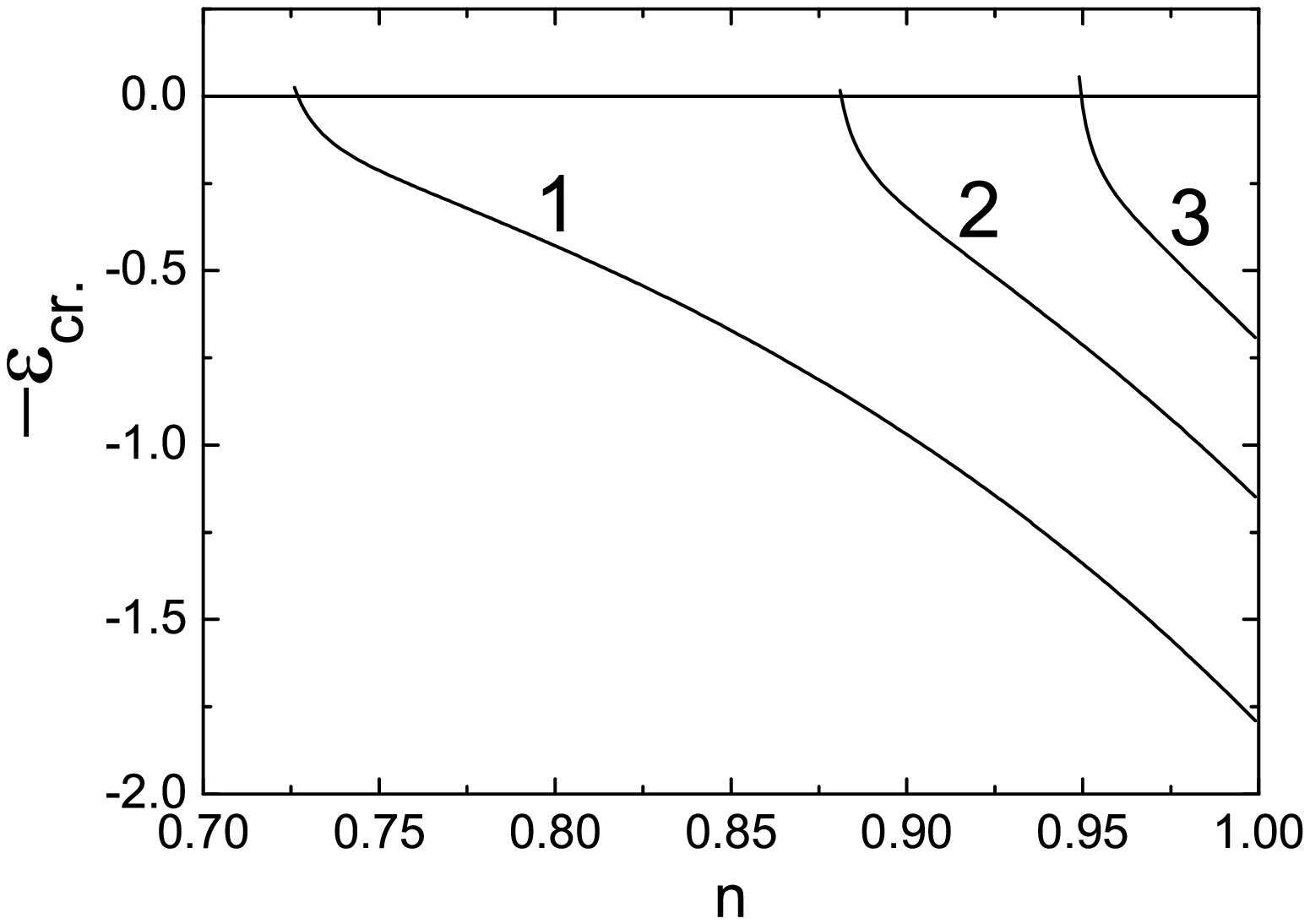}
\end{center}
\caption{The critical values of  -\textit{$\epsilon_{cr.}$}  versus  \textit{n}  from Eq.\eqref{12} on the Fermi level  at $\alpha $=0.1, \textit{g}=0.03, 0.05 and 0.06 (curves 1-3, respectively).}
\label{fig.5}
\end{figure}
\begin{figure}[h]
\begin{center}
\includegraphics[width=\columnwidth]{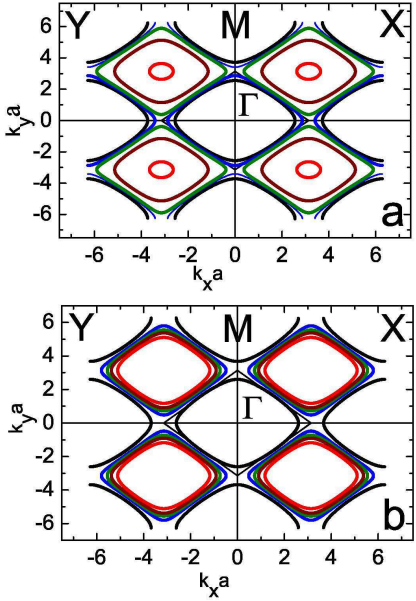}
\end{center}
\caption{ The Fermi surfaces in approximation Hubbard-I square of which is decreased with  increase an electron concentration \textit{n} at $\alpha $=0.1 and   a)  \textit{g}= 0.03 è   \textit{n}=0.727, 0.74, 0.76, 0.85 and 0.99 ($\tilde{\mu }$=0, 0.004,  0.011, 0.057  and 0.171, respectively);  b) \textit{g}=0.06 and \textit{n}=0.95, 0.96, 0.97, 0.98 è 0.999 ($\tilde{\mu }$= 0.0004, 0.0095, 0.0188, 0.0285  and  0.0478, respectively).}
\label{fig.6}
\end{figure}

\section{\label{sec:level1}Effects of d-electron inelastic scattering}

\subsection{\label{sec:level2} Green's function and self-energy}

In the previous section  a detailed consideration of the approximation Hubbard-I was given. Although this approach does not describe an electron scattering it may be a good start to account for  correlation  effects.

To find the time-ordered  Green's function we will consider the contributions of the one-loop diagrams only. It is  the first nonvanishing approximation of  time-dependent perturbation theory with respect to the inverse effective radius of interaction. Since the phonon and  fermion  subsystem are independent, we have 
\[
\begin{array}{l}
 \Lambda _{0\sigma } (p\tau _l ,q\tau _m ) =  -  < T_\tau  \tilde X_p^{0\sigma } (\tau _p )\tilde X_q^{\sigma 0} (\tau _q ) >  \\ 
  =  - U_{ep} (\tau _p  - \tau _q ) < T_\tau  X_p^{0\sigma } (\tau _p )X_q^{\sigma 0} (\tau _q )>  \\ 
 \end{array}
\]
where the first factor    $U_{ep} {\rm (}\tau _{p} -\tau _{q} {\rm )}{\rm}$  = $ {<T}_{\tau } Y_{p}^{} (\tau _{p} )Y_{q}^{+} (\tau _{q} )>_{\hat{\tilde{H}}_{b} }$ is the unperturbed time-ordered uniform bosonic Green's function for the system of Einstein phonons.  Fourier transform of $U_{ep} {\rm (}\tau _{p} -\tau _{q} {\rm )}$  is
\begin{eqnarray}
U_{ep} (i\omega _{n} )=\frac{1}{\beta } \sum _{m=-\infty }^{+\infty }\frac{\psi _{m} }{i\omega _{n} +m\omega _{0} }, 
\label{energy29}
\end{eqnarray}
where $\psi _{m} =2d_{m} \sinh \left(\beta m\omega _{0} /2\right)$, $\omega _{n} =2n{\kern 1pt} {\kern 1pt} \pi /{\kern 1pt} \beta $, $d_{m} =e^{-\lambda ^{2} (2B+1)} I_{m} \left(2\lambda ^{2} \sqrt{B(B+1)} \right)$. ${I_m(x)}$ are the Bessel functions of complex argument, ${\it B=n(}\omega _{0} {\it )}$. ${\it n(x)=1/(exp(}\beta {\it x)-1)}$ is Bose distribution. In the second factor the external Hubbard's operators are not multiplied by bosonic operators $Y_{p}^{} (\tau _{p} )$ but inner ones are unitary transformed. That's why at first  the normalized Green's function is conveniently considered  in form 
\begin{equation}
 \label{30} 
 H_{0\sigma } (p\tau _{{\rm l}} ,q\tau _{{\rm m}} ){\rm =}\frac{\Lambda _{0\sigma } (p\tau _{{\rm l}} ,q\tau _{{\rm m}} )}{U_{ep} {\rm (}\tau _{p} -\tau _{q} {\rm )}}  
\end{equation} 
By $\Sigma _{\sigma } ({\it i}{\kern 1pt} {\kern 1pt} \omega _{n} )$ denote a self-energy part of the total Green's function $\Lambda _{0\sigma } (i\omega _{n} ,{\bf k})$. Then one can write the Dyson's equation for $\Lambda _{0\sigma } (i\omega _{n} ,{\bf k})$ in the next form:
\[\Lambda _{0\sigma } (i\omega _{n} ,{\textbf{k}})=\Sigma _{\sigma } ({\it i}{\kern 1pt} {\kern 1pt} \omega _{n} )+\beta t({\textbf{k}})\Sigma _{\sigma } ({\it i}{\kern 1pt} {\kern 1pt} \omega _{n} )\Lambda _{0\sigma } (i\omega _{n} ,{\textbf{k}})\] 

Apparently, the self-energy is expressed as
\begin{eqnarray}
\Sigma _{\sigma } ({\it i}{\kern 1pt} {\kern 1pt} \omega _{n} )=\sum _{\omega _{n1} }U_{ep} (i\omega _{n} -i\omega _{n_{1} } )H_{0\sigma } (i\omega _{n_{1} } ),
\label{31}
\end{eqnarray} 
where   $H_{0\sigma } (i\omega _{n_{1} } )$ is the Fourier transform of function $H_{0\sigma } (p\tau _{{\rm l}} ,q\tau _{{\rm m}} )$. The poles is supposed to be known. Then it easy to evaluate frequency summation in Eq.\eqref{31} that gives
\begin{eqnarray}
\Sigma _{\sigma } ({\it i}{\kern 1pt} {\kern 1pt} \omega _{n} )=\beta Res\left[f(\omega ){\rm G}_{ph.} {\rm (}i\omega _{n} {\rm -}\omega {\rm )}H_{0\sigma } (\omega )\right]_{{\rm G}_{ph.} } \nonumber \\ +\beta Res\left[f(\omega ){\rm G}_{ph.} {\rm (}i\omega _{n} {\rm -}\omega {\rm )}H_{0\sigma } (\omega )\right]_{H_{0\sigma } } ,
\label{32}
\end{eqnarray}
where the residues are taken in poles of the corresponding Green's functions.

We use the effective self-consistent field in the approximation Hubbard-I  as a start. In this case account must be taken over all the  inner  convolutions  of adjacent unitary transformed Hubbard's operators. For dressed line of  hopping integral $B_{\sigma 0} (\tau _{j} -\tau _{i} ,{\textbf{R}}_{j} -{\textbf{R}}_{i} )$ with Fourier transform  $B_{\sigma 0} (i{\kern 1pt} {\kern 1pt} \omega _{n} ,{\textbf{k}})$ one can write the  graphic equation in form as in Fig.7. Here, the thin and bold wave lines are $\beta t({\textbf{k}})$ and $\beta B_{\sigma 0} (i{\kern 1pt} {\kern 1pt} \omega _{n} ,{\textbf{k}})$, respectively. The thin straight arrow corresponds to  $\tilde{G}_{0\sigma } (i\omega _{n} )<F^{\sigma 0} >$ from Eq.\eqref{08}.
\begin{figure}[h]
\begin{center}
\includegraphics[width=5cm]{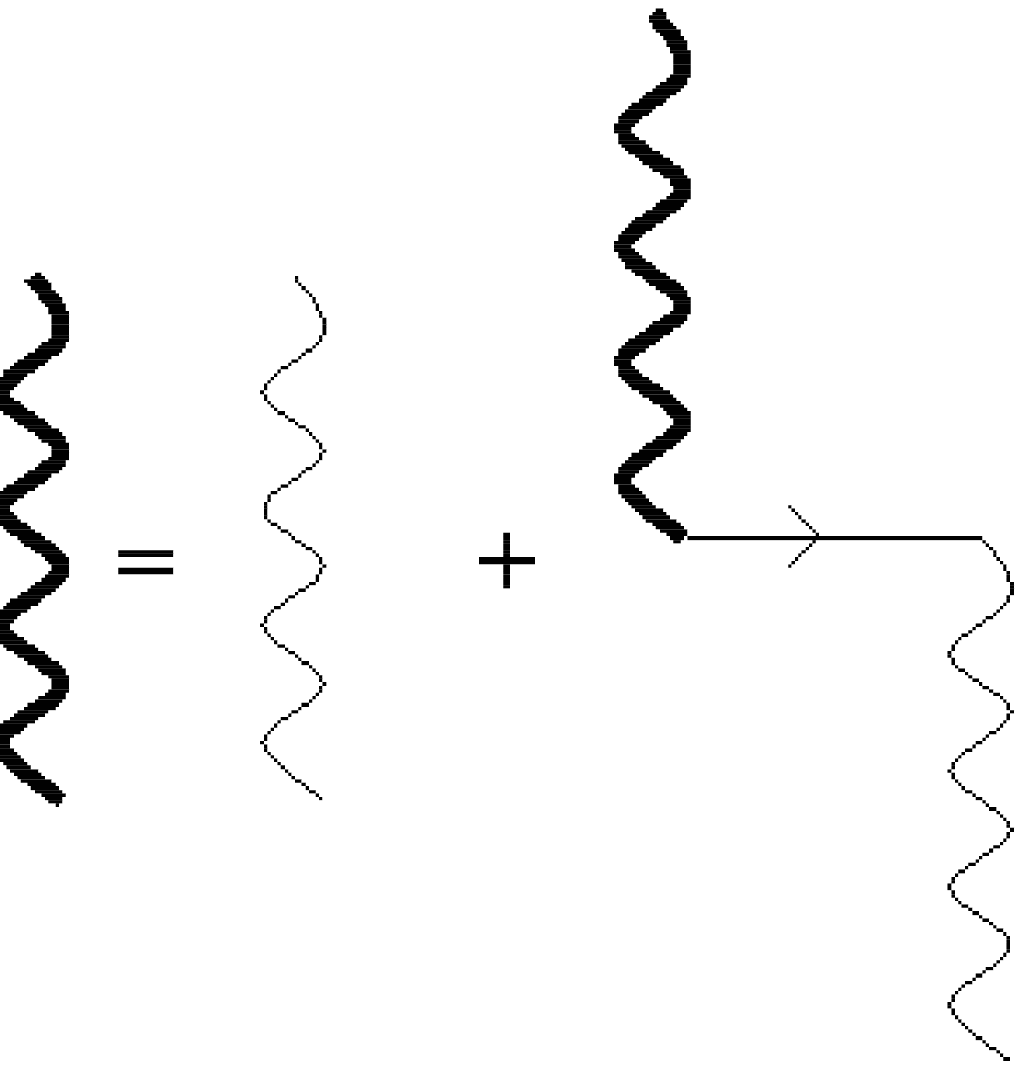}
\end{center}
\caption{ Graphic equation for the effective  hopping integral \textbf{$B_{\sigma 0} (i{\kern 1pt} {\kern 1pt} \omega _{n} ,{\it k})$}.}
\label{fig.7}
\end{figure}
The solution of this equation is trivial and has a form:
\begin{equation} 
\label{33} \beta B_{\sigma 0} (i{\kern 1pt} {\kern 1pt} \omega _{n} ,{\textbf{k}})=\frac{\beta t({\textbf{k}})}{1-\beta t({\textbf{k}})\tilde{G}_{0\sigma } (i\omega _{n} )<F^{\sigma 0} >}  
\end{equation} 
Then in this approach  two main terms must be added to unperturbed Green's function  $G_{0+} (\tau _{p} -\tau _{q} )<F^{+0} >_{0} $:
\[
\begin{array}{l}
 \int\limits_0^\beta  {d\tau _1 \int\limits_0^\beta  {d\tau _2 \frac{1}{N}\sum\limits_{ij\textbf{q}} {\frac{1}{\beta }B_{ + 0} (\tau _1  - \tau _2 ,\textbf{q})e^{i\textbf{q}(\textbf{R}_l  - \textbf{R}_m )} } } } \times \\ 
    < T_\tau  X_p^{0 + } (\tau _p )X_q^{ + 0} (\tau _q )Y_l^ +  (\tau _1 )X_l^{ + 0} (\tau _1 )Y_m^{} (\tau _2 )X_m^{0 + } (\tau _2 ) > _0
 \end{array}
\]
and
\[
\begin{array}{l}
 \int\limits_0^\beta  {d\tau _1 \int\limits_0^\beta  {d\tau _2 \frac{1}{N}\sum\limits_{ij\textbf{q}} {\frac{1}{\beta }B_{ - 0} (\tau _1  - \tau _2 ,\textbf{q})e^{i\textbf{q}(\textbf{R}_l  - \textbf{R}_m )} } } }\times  \\ 
    < T_\tau  X_p^{0 + } (\tau _p )X_q^{ + 0} (\tau _q )Y_l^ +  (\tau _1 )X_l^{ - 0} (\tau _1 )Y_m^{} (\tau _2 )X_m^{0 - } (\tau _2 ) > _0  \\ 
 \end{array}
\]

Using the Vick's theorem for Hubbard's operators and taking into account an independent averaging of  a  product of the  bosonic operators $Y_{m}^{} (\tau _{2} )$ we evaluate these integrals. In Fig.8 the self-energy terms in form to linked diagrams ( points in Fig.8 denote index convolutions) are presented generally for $\sigma $=$\pm$ . In figure the dash line with arrow corresponds to bosonic unperturbed Green's function $G_{+-} (i\omega _{n} )=1/(\beta (i\omega _{n} +\varepsilon _{+} -\varepsilon _{-} ))$ where $\varepsilon _{+} -\varepsilon _{-} =-h$ and \textit{h} is the effective exchange field.
\begin{figure}[h]
\begin{center}
\includegraphics[width=\columnwidth]{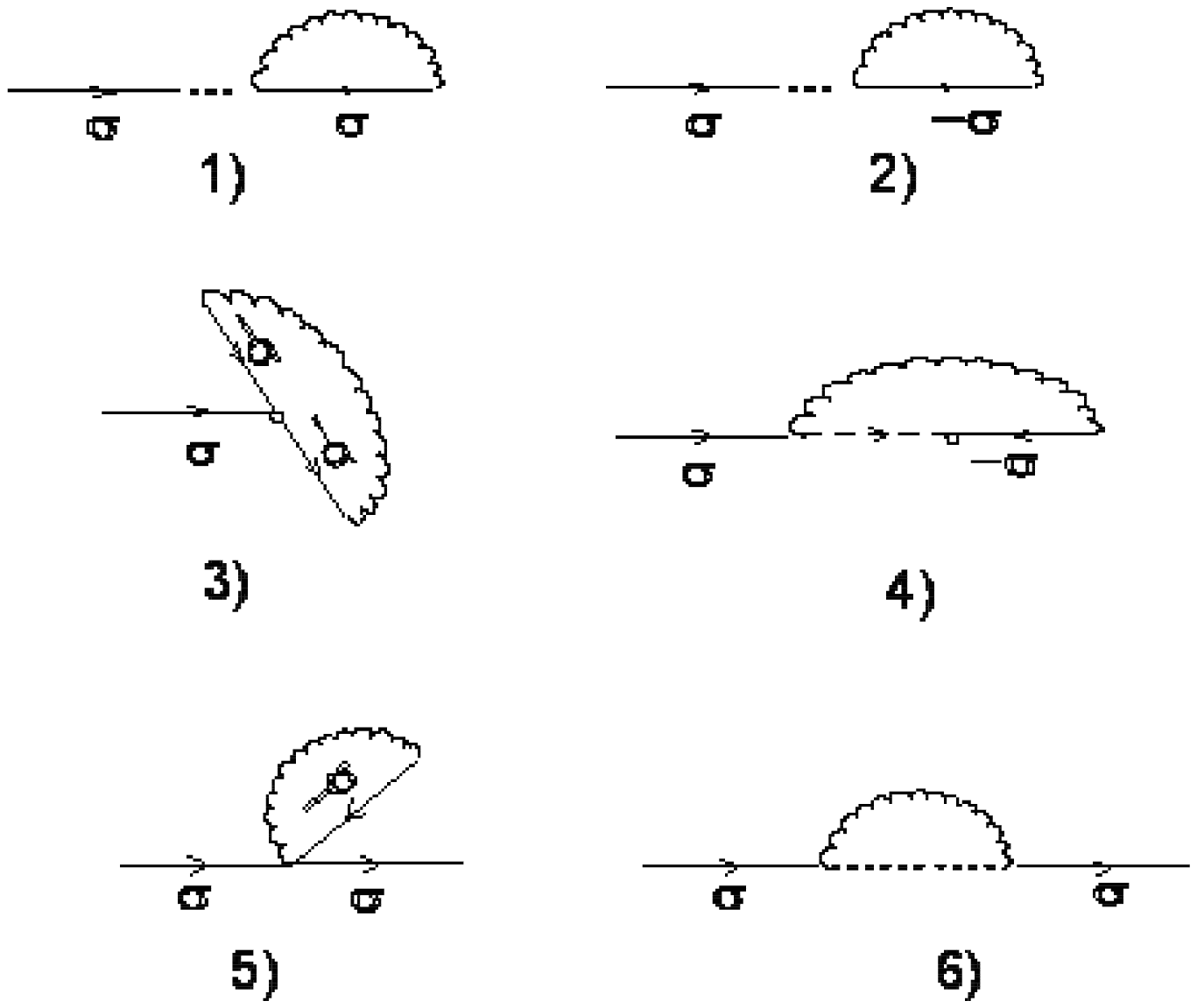}
\end{center}
\caption{Diagrams for the self-energy $\Sigma _{\sigma } ({\it i}{\kern 1pt} {\kern 1pt} \omega _{n} )$ in the first approximation of the perturbation theory \cite{14}.}
\label{fig.8}
\end{figure}

Let us write the analytic expressions for diagrams 1-3:
\[
\begin{array}{l}
 {\rm 1)} - \beta \delta \tilde \mu _\sigma ^{} \partial _\sigma   < F^{\sigma 0}  > _0 G_{0\sigma } (i\omega _n ) =  - \frac{1}{N}\sum\limits_{\textbf{q}n_1 } {{\kern 1pt} \beta }  \\ 
  \times \tilde G_{0\sigma } (i\omega _{n_1 } )B_{\sigma 0} (i\omega _{n_1 } ,{\bf q})\partial _\sigma   < F^{\sigma 0}  > _0 G_{0\sigma } (i\omega _n ) \\ 
 2) - \beta \delta \tilde \mu _{ - \sigma }^{} \partial _{ - \sigma }  < F^{\sigma 0}  > _0 G_{0\sigma } (i\omega _n ) =  - \frac{1}{N}\sum\limits_{\textbf{q}n_1 } {{\kern 1pt} \beta }  \\ 
  \times \tilde G_{0 - \sigma } (i\omega _{n_1 } )B_{ - \sigma 0} (i\omega _{n_1 } ,{\bf q})\partial _{ - \sigma }  < F^{\sigma 0}  > _0 G_{0\sigma } (i\omega _n ) \\ 
 3)\;\tilde \nu _{ - \sigma }  < F^{ - \sigma 0}  > _0 G_{0\sigma } (i\omega _n )\, = \frac{1}{N}\sum\limits_{\textbf{q}n_1 m_1 } {B_{ - \sigma 0} (i\omega _{n_1 } ,{\bf q}){\kern 1pt} }  \\ 
  \times U_{ep} (i\omega _{n_1 }  - i\omega _{m_1 } )\left[ {G_{0 - \sigma } (i\omega _{m_1 } )} \right]^2  < F^{\sigma 0} > _0 \beta G_{0\sigma } (i\omega _n ) \\ 
 \end{array}
\]
Here,  the derivative  $\partial _{\sigma } <F^{\sigma 0} >_{0} =\partial <F^{\sigma 0} >_{0} /\partial (-\beta {\kern 1pt} \varepsilon _{\sigma } )$ and ${\kern 1pt} \varepsilon _{\sigma } =-\tilde{\mu }$. These diagrams are proportional to external unperturbed Green's function $G_{0\sigma } (i\omega _{n} )=1/(\beta (i\omega _{n} +\tilde{\mu }))$. One can factor out and add $G_{0\sigma } (i{\kern 1pt} {\kern 1pt} \omega )<F^{+0} >_{0} $. Then in parentheses we obtain a series expansion for an average $<F^{\sigma 0} >=1-n/2$. This series was considered in work [15]. It  gives an equation for chemical potential $\tilde{\mu }$. Therefore, the sum of diagram terms 1-3 and  $  G_{0\sigma } (i{\kern 1pt} {\kern 1pt} \omega )<F^{+0} >_{0} $ gives  a contribution $G_{0\sigma } (i{\kern 1pt} {\kern 1pt} \omega _{n} )(1-n/2)$ to $H_{0\sigma } (i\omega _{n_{1} } )$.

The contribution from diagram 5  to $H_{0\sigma } (i\omega _{n_{1} } )$ is written as
5)  $-\beta \delta \tilde{\mu }_{-\sigma }^{} <F^{\sigma 0} >_{0} G_{0\sigma }^{2} (i\omega _{n} )$
One can prove that diagrams 4 and 6 only  differs by factors $<X^{\sigma \sigma }>_{0}$ and $ <F^{\sigma 0}>_{0} / (1-<F^{\sigma 0}>_{0})$, respectively, i.e. one writes 
\[
\begin{array}{l}
4)+6)= K_{0} (I_{1} (i\omega _{n} )+I_{2} (i\omega _{n}))G_{0\sigma }^{2} (i{\kern 1pt} {\kern 1pt} \omega _{n} ), 
 \end{array}
\]
where $K_{0} =<X^{\sigma \, \sigma }>_{0} +<F^{\sigma 0}>_{0} (1-<F^{\sigma 0}>_{0} )$=3/4 at \textit{T}=0. Evaluating the sum over inner discrete frequencies  by means  of previously considered method of inverse function we write the final expression for $H_{0\sigma} (i\omega _{n})$:
\begin{eqnarray}
 H_{0\sigma } (i\omega _n ) = G_{0\sigma } (i{\kern 1pt} {\kern 1pt} \omega _n )(1 - n/2) + K_0 (I_1 (i\omega _n ) \nonumber \\ 
  + I_2 (i\omega _n ))G_{0\sigma }^2 (i{\kern 1pt} {\kern 1pt} \omega _n ) - \beta \delta \tilde \mu _{ - \sigma }^{}  < F^{\sigma 0}  > _0 G_{0\sigma }^2 (i\omega _n ), 
\label{34}
\end{eqnarray} 
\begin{equation}
\label{35} 
\begin{array}{l} {I_{1} (i\omega _{n} )
=\int _{-\infty }^{+\infty }d\Omega \frac{4\beta f(\Omega )\tilde D_C \left( {\Omega ,\ \ \alpha } \  \right)}{\left[F(\Omega )<F^{\sigma 0} >\right]^{2} } \beta U_{ep} (\Omega -i\omega _{n} ){\kern 1pt}  } \\ \\ {I_{2} (i\omega _{n} )=\frac{1}{N} \sum\limits_{{\textbf{q}}m_{1} =-\infty }^{+\infty }\frac{\beta t({\textbf{q}})({\it n(m}_{{\it 1}} {\it \omega }_{{\it 0}} {\it )}+1)\psi _{m_{1} } }{1-t({\textbf{q}})<F^{\sigma 0} >\beta \tilde{G}_{0\sigma } (i\omega _{n} -{\it m}_{{\it 1}} {\it \omega }_{{\it 0}} )}  } 
\end{array}  
\end{equation}
where $\tilde D_C (\Omega ,\alpha )=D_C \left( {\frac{4}{{F(\Omega ) < F^{\sigma 0}  > }},\alpha } \right)$.
It follows from presented above expressions that diagrams 4 and 6 describe electron-polaron and electron-electron scattering only. The rest of diagrams renormalize the excitation spectrum. At T ${\sim}$ 0 the factor $({\it n(m}_{{\it 1}} {\it \omega }_{{\it 0}} {\it )}+1)\psi _{m_{1} } $in $I_{2} (i\omega _{n} )$ may be simplified. It easy to find that $\psi _{m} \approx \tilde{\psi }(\left|m\right|)sign(m)\; $, where $\tilde{\psi }(m)=e^{-\lambda ^{2} } \frac{\lambda ^{2m} }{m!} \; ,\; sign(0)=0$ and $({\it n(m}_{{\it 1}} {\it \omega }_{{\it 0}} {\it )}+1)\psi _{m_{1} } \approx \tilde{\psi }(\left|m\right|)\theta (m)$. After analytic continuation $i{\kern 1pt} {\kern 1pt} \omega _{n} \to \omega +i\delta $ the real part of integral $I_{1} (\omega )$ is evaluated  numerically as Cauchy  principal  value. The imaginary part of  $I_{1} (\omega )$ is easy calculated since $I_m[\beta U_{ep} (\Omega -\omega -i\delta )]$ gives the Dirac delta function.

To obtain the final expression for self-energy $\Sigma _{\sigma } ({\it i}{\kern 1pt} {\kern 1pt} \omega _{n} )$ it is necessary to substitute $H_{0\sigma } (i\omega _{n} )$ in Eq.\eqref{32}. As result in the limit T ${\sim}$ 0  we have
\begin{eqnarray} 
\label{36} 
\Sigma _{\sigma } ({\it i}{\kern 1pt} {\kern 1pt} \omega _{n} )=P_{0} ({\it i}{\kern 1pt} {\kern 1pt} \omega _{n} )+P_{5} ({\it i}{\kern 1pt} {\kern 1pt} \omega _{n} )+P_{46} ({\it i}{\kern 1pt} {\kern 1pt} \omega _{n} )
\end{eqnarray} 
where
\begin{equation} 
\begin{array}{l} 
{P_{0} (\omega )=\frac{1-0.5n}{\beta (\omega +\tilde{\mu })} M\left(1,1+\frac{\omega +\tilde{\mu }}{\omega _{0} } ,-\lambda ^{2} \right)} \\ {P_{5} (\omega )=-<F^{\sigma 0} >_{0} \frac{1}{\beta } \delta \mu _{-\sigma } \sum\limits_{m=0}^{\infty }e^{-\lambda ^{2} } \frac{\lambda ^{2m} }{m!} \frac{1}{(\omega +m\omega _{0} +\tilde{\mu })^{2} }} \\  =-\frac{1}{\beta } \frac{J_{5} (\omega )}{(\omega +\tilde{\mu })^{2}} \\ 
{P_{46} (\omega )=J_{1} (\omega )+J_{2} (\omega )+J_{3} (\omega )}
\label{37}
\end{array} 
\end{equation} 
Here, ${\it w(}\omega {\it )=}\frac{\omega +\tilde{\mu }}{\omega _{0} } $  and  $J_{5}(\omega )={<F^{\sigma 0} >_{0}} \delta \mu _{-\sigma } e^{-\lambda ^{2} }{}_{2} F{}_{2} ({\it w},{\it w};1+{\it w},1+{\it w};\lambda ^{2} )$,  where ${}_{2} F{}_{2} ({\it w},{\it w};1+{\it w},1+{\it w};\lambda ^{2} )$ is the generalized  hypergeometric function. For  T=0  we have $<F^{\sigma 0} >_{0} =1/2$  and
\begin{eqnarray}
\begin{array}l 
J_{1} (\omega )=\sum _{m_{1} ,m_{2} =-\infty }^{+\infty }K_{0} \frac{1}{\beta ^{2} } \frac{\psi _{m_{1} } \psi _{m_{2} } ({\it n}(m_{1} \omega _{0} )-{\it n}(m_{2} \omega _{0} ))}{(\omega +{\it i}\delta +m_{1} \omega _{0} +\tilde{\mu })^{2} }  \\ \times\int _{-\infty }^{+\infty }d\Omega   \frac{{\rm Z} (\Omega ,n,\alpha )f(\Omega )}{\Omega -\omega -i\delta +(m_{2} -m_{1} )\omega _{0} }, \kern 30 pt 
\label{38}
\end{array}
\end{eqnarray}
where
 ${\rm Z}(\Omega ,n,\alpha ) = \tilde D_C \left( \Omega  ,-\alpha \right) \frac{{4\beta }}{{\left( {\beta \tilde G_{0\sigma } (\Omega ) < F^{\sigma 0}  > } \right)^2 }}$
\begin{widetext}
\begin{eqnarray}
\label{39} 
J_{2} (\omega )=\sum _{m_{1} ,m_{2} =-\infty }^{+\infty }K_{0} \frac{1}{\beta ^{2} } \frac{\psi _{m_{1} } \psi _{m_{2} } ({\it n}(m_{1} \omega _{0} )-n(m_{2} \omega _{0} ))}{(\omega +{\it i}\delta +m_{1} \omega _{0} +\tilde{\mu })^{2} } \frac{1}{N} \sum _{{\textbf{q}}}\frac{\beta t({\textbf{q}})({\it n}(\left[m_{2} -m_{1} \right]\omega _{0} )+1)}{1-t({\textbf{q}})<F^{\sigma 0} >\beta \tilde{G}_{0\sigma } (\omega +{\it i}\delta -\left[m_{2} -m_{1} \right]\omega _{0} )}
\end{eqnarray}
\begin{eqnarray} 
\label{40} 
J_{3} (\omega )=\sum _{m_{1} ,m_{2} =-\infty }^{+\infty }K_{0} \frac{1}{\beta ^{2} } \frac{\psi _{m_{1} } \psi _{m_{2} } ({\it n}(m_{2} \omega _{0} )+f(-\tilde{\mu }))}{(\omega +{\it i}\delta +m_{1} \omega _{0} +\tilde{\mu })^{2} } \int _{-\infty }^{+\infty }d\Omega   \frac{{\rm Z} (\Omega ,n,\alpha )f(\Omega )}{\Omega +m_{2} \omega _{0} +\tilde{\mu }} \cdot \left(1+\frac{\omega +m_{1} \omega _{0} +\tilde{\mu }}{\Omega +m_{2} \omega _{0} +\tilde{\mu }} \right) 
\end{eqnarray} 
\end{widetext}
\begin{eqnarray} 
\label{41} 
\delta \mu _{-\sigma } =\int _{-\infty }^{+\infty }\frac{1}{\beta } {\rm Z} (\Omega ,n,\alpha )\beta \tilde{G}(\Omega )f(\Omega )
\end{eqnarray}
In Eq.\eqref{39} for ${m_1=m_2}$  the uncertainty is observed. Thus, in the limit ${m_1}\to{m_2}$ we have the function
\begin{eqnarray} 
 J_0 (\omega ) = \frac{{K_0e^{ - 2\lambda ^2 } }}{{\beta (\omega  + i\delta  + \tilde \mu )^2 }} \kern 100pt \nonumber \\ 
  \times \frac{1}{N}\sum\limits_{\textbf{q} =  - \infty }^{ + \infty } {\frac{1}{{1 - t(\textbf{q}) < F^{\sigma 0}  > \beta \tilde G_{0\sigma } (\omega  + i\delta )}}}
\label{42}
\end{eqnarray}
which is determined by lattice Green's function $G(s,\alpha )$. And so, for sum \eqref{39} the term with ${m_1=m_2}$ is excluded. It is necessary to add the expression \eqref{42} to Eq.\eqref{39}  instead  of  this abnormal term.

 One writes for  imaginary part $J_{1} (\omega )$ at {T=0}
\begin{eqnarray}
ImJ_{1} (\omega )=\frac{4\pi K_{0} }{\beta \left(1-0.5n\right)^{2} } \kern 100pt \nonumber \\
\times \sum _{m=-\infty }^{\infty }R(\omega ,\tilde{\mu },m)\omega _{0}^{2} \varphi \left(\frac{\omega +\tilde{\mu }}{\omega _{0} } ,n\right)\theta (m\omega _{0} -\omega ),       
\label{43}  
\end{eqnarray} 
where
\begin{eqnarray} 
\begin{array}l
\label{44} 
\varphi (w,n)=\tilde D_C \left( w\Omega_0-\mu,-\alpha \right)\frac{w^{2} }{M^{2} (1,1+w,-\lambda ^{2} )}
\end{array}
\end{eqnarray} 
\begin{widetext}
\[R(\omega ,\tilde{\mu },m)=\sum _{m_{1} =-\infty }^{+\infty }\frac{\left\{\tilde{\psi }(\left|m+m_{1} \right|)\tilde{\psi }(\left|m_{1} \right|)sign(m+m_{1} )\theta (m_{1} )-\tilde{\psi }(\left|m_{1} \right|)\tilde{\psi }(m+\left|m_{1} \right|)sign(m_{1} )\theta (m+m_{1} )\right\}}{(\omega +\tilde{\mu }+m_{1} \omega _{0} )^{2} }  \] 
Notice that $R(\omega ,\tilde{\mu },m)=0$ at $\textit{m}$=0. The real part  of  $J_{1} (\omega )$ and complex function $J_{2} (\omega +i\delta )$ are expressed as 
\begin{eqnarray} 
\label{45} 
ReJ_{1} (\omega )=\frac{1}{\beta } K_{0} \sum _{m=0}^{\infty }R(\omega ,\tilde{\mu },m)I_{11} (\omega ,n,m) \\ 
J_{2} (\omega )=\frac{K_{0} }{\beta } \sum _{m=-\infty }^{+\infty }R(\omega ,n,m) S_{2} (\omega ,n,\alpha ,m)
\end{eqnarray}
Here,
\begin{eqnarray}
\label{47}
I_{11} (\omega ,n,m)=\frac{{\kern 1pt} 4\omega _{0}^{2} }{(1-0.5{\it n})^{2} } \left\{\sum _{k=-\infty }^{-1}V.p.\int _{k}^{k+1}dw\frac{\varphi (w)}{w-\left(\frac{\omega +\tilde{\mu }}{\omega _{0} } +m\right)} +V.p.\int _{0}^{\tilde{\mu }/\omega _{0} }dw\frac{\varphi (w)}{w-\left(\frac{\omega +\tilde{\mu }}{\omega _{0} } +m\right)}    \right\}
\end{eqnarray}
\begin{eqnarray}
\label{48}
S_{2} (\omega ,n,\alpha ,m)=\frac{\omega +\tilde{\mu }-m\omega _{0} }{\left(1-0.5n\right)M\left(1,1+\frac{\omega +\tilde{\mu }}{\omega _{0} } -m,-\lambda ^{2} \right)} \left\{-1+(\omega +\tilde{\mu }-m\omega _{0} )S_{3} (\omega ,n,\alpha ,m)\right\}
\end{eqnarray}
\begin{eqnarray}
\label{49}
S_{3} (\omega ,n,\alpha ,m)=\frac{4}{\left(1-0.5n\right)M\left(1,1+\frac{\omega +\tilde{\mu }}{\omega _{0} } -m,-\lambda ^{2} \right)} G^{*} \left(\frac{4\omega _{0} \left[\frac{\omega +\tilde{\mu }}{\omega _{0} } -m\right]}{(1-0.5n)M\left(1,1+\frac{\omega +\tilde{\mu }}{\omega _{0} } -m,-\lambda ^{2} \right)} ,\alpha \right),
\end{eqnarray}
\end{widetext}
where  the integrals  are evaluated  numerically as Cauchy  principal  value and integrand $\varphi (w,n)$ is determined by Eq.\eqref{44}. $G^ *  (s,\alpha )$ is the conjugate lattice Green's function. 

Thus, Eqs.\eqref{36}-\eqref{49} contains  a total information about spectral properties of the cuprate d-electron subsystem with electron-phonon binding. Below it is given a  numerical solution  of  the previously obtained  equations to analyze both the excitation spectrum  and dissipation  were caused by electron-electron and electron-polaron interactions. Finally, let us write the contribution  $\textit{P}_{46}$($\omega$) in form:
\[P_{46} (\omega )=J_{0} (\omega )+J_{1} (\omega )+J_{2} (\omega )+J_{3} (\omega ),\] 
where $R(\omega ,\tilde{\mu },m)=0$ for $\textit{m}=0$.
\subsection{\label{sec:level2} Electron spectrum in the absence of electron-phonon interaction}

In this section we will study  the dynamical properties of electrons for temperature $T$=0  in the absence of electron phonon binding, i.e. at  $g$=0. Before we proceed further, we evaluate numerically the chemical potential  $\tilde{\mu }$. In works \cite{15,23} it was presented an approach to calculate the chemical potential  versus  band filling. In particular, in \cite{15}  we obtained Eq.(51) for $\tilde{\mu }$ versus \textit{n}  in  PM-2  state  with the nearest hopping energy. One can generalize this equation when the influence of  next-nearest neighbors is taken into account, i.e. for $\alpha$$\neq$0:
\begin{eqnarray}
\tilde \mu  = \frac{{2 - n}}{8}I^{ - 1} \left( {1 - \frac{1}{2}(1 - n)(2 - n),\,{\kern 1pt} \alpha } \right)
\label{50}
\end{eqnarray}
were $I^{ - 1} (x,\alpha )$ is the inverse function of function $I(x,\alpha )$ (see Eq. \eqref{28}). Similarly, the term  $\delta \mu _{\sigma } $ of diagram 5  in Fig.8  is 
\begin{eqnarray}
\label{51}
\delta \mu _{\sigma } =E\left(I^{-1} \left(1-\frac{1}{2} (1-n)(2-n),\, {\kern 1pt} \alpha \right)\right),
\end{eqnarray}
where
\begin{eqnarray} 
\label{52} 
E(x,\alpha ) = \int_{ - 2 - 2\alpha }^x {xD_C (x,{\kern 1pt} {\kern 1pt} {\kern 1pt} \alpha )dx}
\end{eqnarray}

In Fig.9 a  the concentration dependencies of chemical potential at $\alpha$= -0.4, -0.3, -0.2, -0.1, 0, 0.1, 0.2, 0.3 è 0.4 (curves 1-9, respectively) are presented. In Fig.9 b an area of   existence of the PM-2 phase in coordinates $\alpha$-\textit{n } is shown. From Eq.\eqref{50} it  follows that chemical potential $\tilde{\mu }$ of filled band  coincides with the solution \eqref{28}  in approximation Hubbard-I.
\begin{figure}[h]
\begin{center}
\includegraphics[width=\columnwidth]{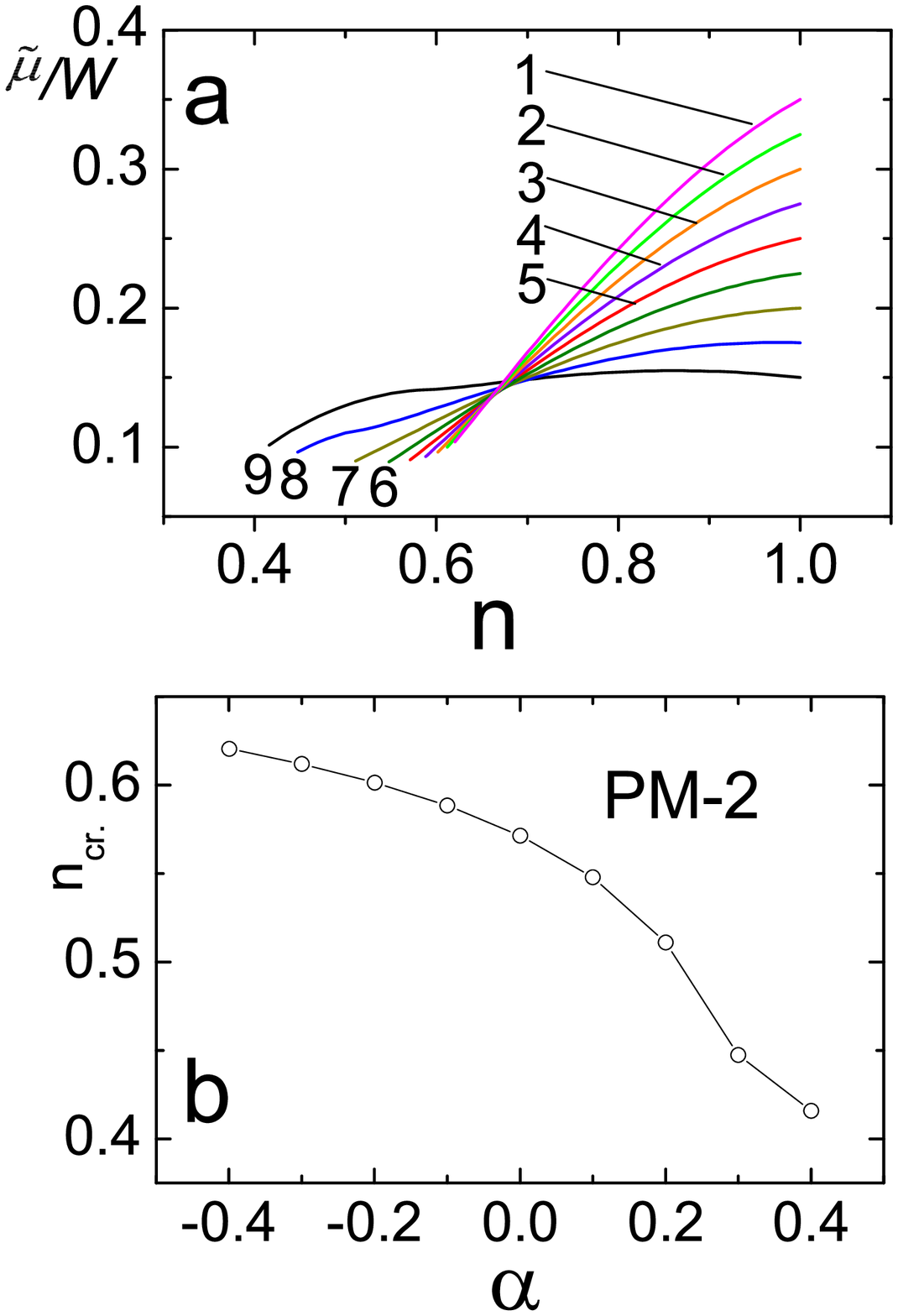}
\end{center}
\caption{ a) The concentration dependencies of the chemical potential in PM-2 phase  with \textit{g}=0  and  $\alpha$= -0.4, -0.3, -0.2, -0.1, 0, 0.1, 0.2, 0.3 è 0.4 (curves 1-9, respectively);  b)  phase diagram in the $\alpha$-\textit{n}  plane  at \textit{T}=0.}
\label{fig.9}
\end{figure}

The spectral density function is expressed as 
\begin{widetext}
\begin{eqnarray}
\label{53} 
A({\it k},\, \, \omega )=-\frac{1}{\pi } \frac{(\omega +\tilde{\mu })^{4} \beta Im\Sigma (\omega )}{\left[(\omega +\tilde{\mu })^{2} -t({\it k})(\omega +\tilde{\mu })^{2} \beta Re\Sigma (\omega )\right]^{2} +(\omega +\tilde{\mu })^{2} t^{2} ({\it k})\left[(\omega +\tilde{\mu })^{2} \beta Im\Sigma (\omega )\right]^{2} }  
\end{eqnarray}
In view of \eqref{32} ${\rm}$ at ${T = 0}$ and ${K_{\rm 0} {\rm  = 3/4 }}$ we have 
\begin{eqnarray}
\label{54}
\begin{array}l
{(\omega +\tilde{\mu })^{2} \beta Re\Sigma (\omega )=(\omega +\tilde{\mu })\left[1-0.5n-\frac{3}{4(1-0.5n)} \right]+\frac{3(\omega +\tilde{\mu })^{2} }{(1-0.5n)^{2} } ReG\left(\frac{4(\omega +\tilde{\mu })}{1-0.5n} ,-\alpha \right)-0.5\, \delta \mu _{\sigma } }
(\omega +\tilde{\mu })^{2} \\
\beta Im\Sigma (\omega )=-\frac{3(\omega +\tilde{\mu })^{2} }{(1-0.5n)^{2} } \pi D_{C} \left(\frac{4(\omega +\tilde{\mu })}{1-0.5n} ,-\alpha \right)
\end{array}
\end{eqnarray}
\end{widetext}
Taking into account Eq.\eqref{54} it easy to write the dispersion equation 
\begin{eqnarray} 
\begin{array}l
\label{55}
- \varepsilon _{c.} (\Omega _k ) \kern 100 pt  \\
= \cos (k_x a) + \cos (k_x a) + 2\alpha \cos (k_x a)\cos (k_x a) \\ 
=-\frac{{4(\Omega _k  + \tilde \mu )^2 }}{{(\Omega _k  + \tilde \mu )^2 \beta {\mathop{\rm Re}\nolimits} \Sigma (\Omega _k )}}
\end{array}
\end{eqnarray} 
From Eq.\eqref{55} it is seen that now the excitation spectrum depends on real part  $ReG(s,-\alpha )$ of the lattice Green's function \eqref{15} which has both  two singularities at the edges of unperturbed electron band, i.e. at ${s{\rm  = }s_{\rm L} {\rm   =   - 2(1 + }\alpha {\rm  ) }}$ and  ${s{\rm  = }s_{\rm R} {\rm   =  2(1 -  }\alpha {\rm )}}$. Also, there is  one  breakdown at   \textit{s}=0 (see Fig.3 a) that corresponds to frequency ${\Omega _{{\it k}} =-\tilde{\mu }}$. These circumstances cause the cardinal changes to the spectrum of  the electron excitations in comparison with approximation   Hubbard-I.

In Fig. 10  the resonance frequencies   $\Omega _{{\it k}} $  along the  nodal direction for bismuth cuprate  2212 at different electron band filling are shown. In Fig.10  the values of  \textit{n}  correspond  to the most  typical  points of  function $-\varepsilon _{c.} $(\textit{n})  from \eqref{55} at $\Omega _{{\it k}} =0$ (Fermi level)  to be shown in insert to figure. So, at $\alpha$=0 (curve 1 in insert ) the curve  $-\varepsilon _{c.} $(\textit{n)}  tends to own lower edge -2 at \textit{n}=0.7. For this value of  \textit{n}  a small gap arises on  the Fermi level. It  has  a  correlation origin that is reflected in Fig.10 a (curve 2).  A such gap (see Fig.10,  curves 1 and 3)  is absent  at other values of \textit{n}. The function $-\varepsilon _{c.} (n)$ does not approach the edge value -2-2$\alpha $ with  increase  an  influence of  the next-nearest neighbours where paramagnetic phase is realized. In this case a gap does  not  appear on the Fermi level for all possible values of \textit{n }(see Fig. 10 b). One can point out the next main peculiarities of the spectrum in Fig.10 b. The kinks in  the dispersion $\Omega _{{\it k}} $ at  $\Omega _{\rm k}  = \omega _L  =  - 0.531$ and $\Omega _{\rm k}  = \omega _R  = 0.019$ for n=0.9 (curve 3) are determined by equation
\begin{eqnarray} 
\label{56} \frac{4(\omega +\tilde{\mu })}{1-0.5n} -s=0 
\end{eqnarray} 
at  ${\rm  }s = s_L$ and  $s_R$, respectively. From this equation we obtain the lower $\omega _{L} =-\tilde{\mu }-0.5(1+\alpha )(1-0.5n)$ and upper $\omega _{L} =-\tilde{\mu }+0.5(1-\alpha )(1-0.5n)$ edges of incoherent spectrum the bandwidth $\Delta \omega _{LR} =1-0.5n$ of which is determined by  electron concentration only.
\begin{figure}[h]
\begin{center}
\includegraphics[width=\columnwidth]{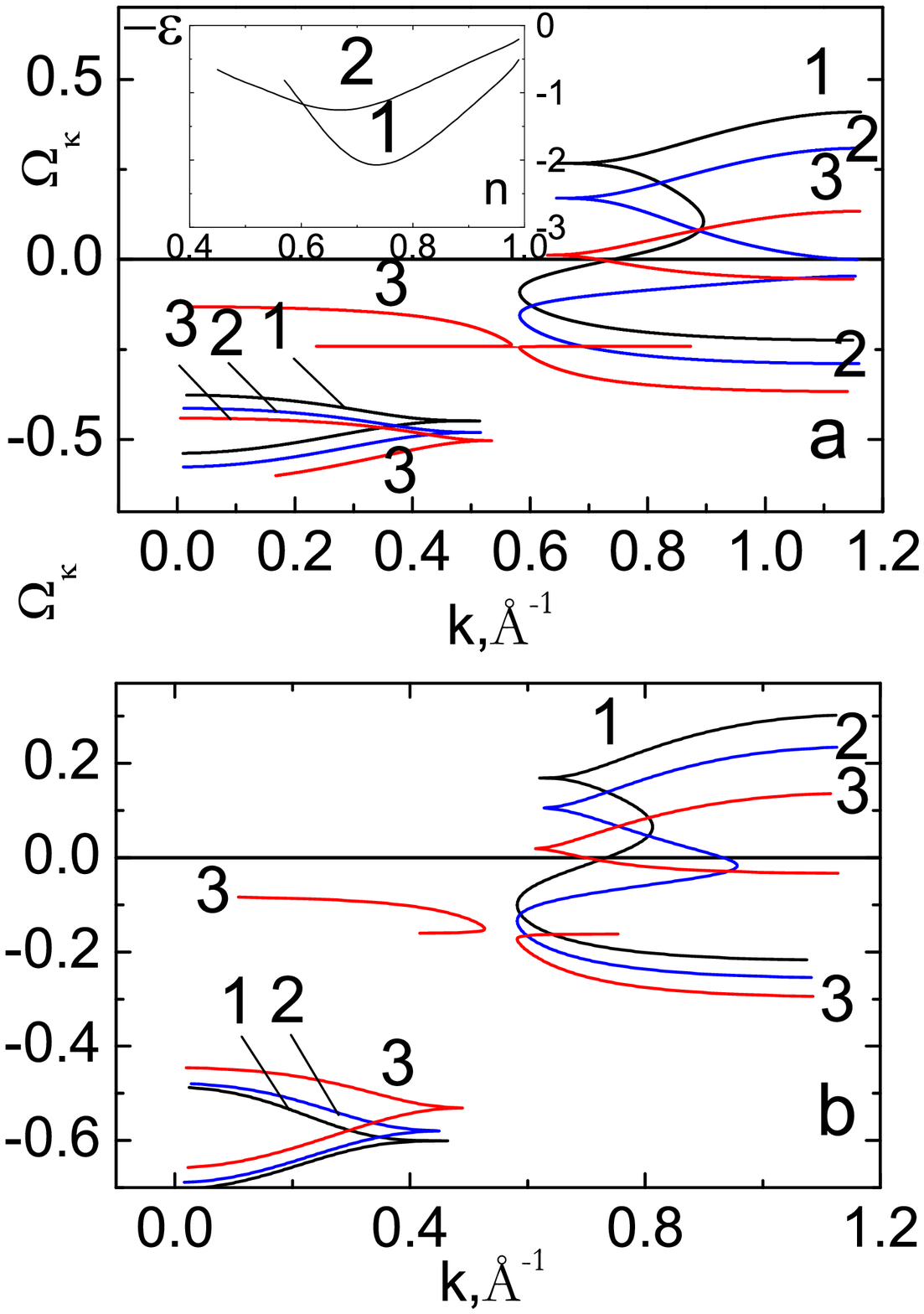}
\end{center}
\caption{The electron excitation frequency-momentum dispersions for bismuth cuprate 2212   (lattice constant {$a=3.814 \AA$}) along the nodal direction without electron-phonon binding: a) $\alpha $=0,\textit{ n}=0.571 ($\tilde{\mu }$=0.090), \textit{n}=0.7 ($\tilde{\mu }$=0.155) and  \textit{n}=0.97 ($\tilde{\mu }$=0.245);  b) $\alpha $=0.3, \textit{n}=0.46 ($\tilde{\mu }$=0.101), \textit{n}=0.63 ($\tilde{\mu }$=0.134) è  \textit{n}=0.90 ($\tilde{\mu }$=0.173) (curves 1-3, respectively). In inset: -$\varepsilon _{c.} $ versus \textit{n} on the Fermi level ($\Omega _{{\it k}} $=0)  from dispersion Eq. \eqref{55} for $\alpha $=0 and $\alpha $=0.3  (curves 1 and  2,  respectively).}
\label{fig.10}
\end{figure}
Thus, the imaginary part of the lattice Green's function  is equal zero out of  indicated area  and  the electron excitations are coherent. Also, two peculiarities are determined by equation \eqref{56} at \textit{s}=0 and  \textit{s}=2$\alpha$. In this case the real and imaginary parts of $G(s,\alpha )$ have discontinuity and van Hove singularity, respectively. In excitation spectrum near the indicated points there is  a sufficiently complicated dispersion with appearance of  pseudogaps.

Therefore, the inclusion of the electron-electron scattering in cuprate  planes causes an essential rebuilding of the excitation spectrum in comparison with approximation Hubbard-I where this scattering is absent. A correlation gap appears on the Fermi level in the area of electron concentrations when $-\varepsilon _{c.} $(0)$\leq -2+2\alpha$  that leads to disappearance of the Fermi surface.

In Fig.11 the Fermi surfaces are presented  with account for electron-electron scattering in accordance with formula \eqref{55} at  $\Omega _{{\it k}} =0$. The no-monotone change in Fermi surface  square in Fig.11 a and b takes place in accordance with dependences  $-\varepsilon _{c.} $(\textit{n}) ( see curves 1 and 2 in inset to Fig. 10 a). In Fig.11 b the  electron Fermi surfaces at  \textit{n}=0.46 and 0.63 are shown. At \textit{n}=0.9 we have the hole Fermi surface. 
\begin{figure}[h]
\begin{center}
\includegraphics[width=\columnwidth]{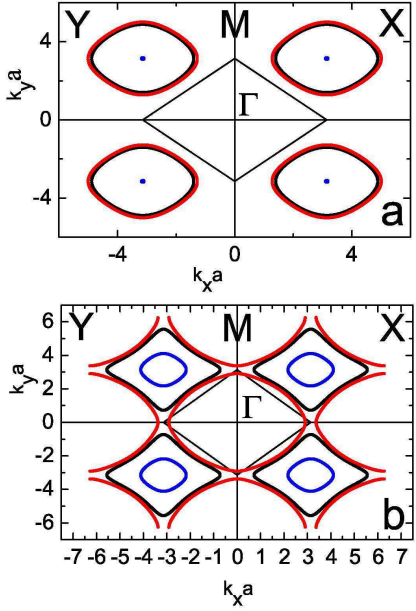}
\end{center}
\caption{The electron and hole Fermi surfaces in the absence of the electron-phonon binding. The values of the parameters are as in Fig.10.  In  Fig. 11 a  the inner surfaces as point and big oval are obtained for  \textit{n}=0.7 and  0.571, respectively. The outside surface corresponds to \textit{n}=0.97.  In Fig. 11 b the electron Fermi surfaces as deformed square  and inner oval centered out of $\Gamma$ point are shown for \textit{n}=0.46 and 0.63, respectively. The hole Fermi surface has a shape of the big oval centered on point $\Gamma$ at \textit{n}=0.9, where $\Gamma$ is the center of  Brillouin band.}
\label{fig.11}
\end{figure}

It is interesting  to study the corresponding spectral densities using Eq.\eqref{54}. In accordance with ARPES terminology \cite{1}  let EDC  denote the frequency dependence $A({\textbf{k}},\,\omega )$ along  nodal direction  (${k_x=k_y}$) of ${\textbf{k}}$ at fixed value of its module. MDC spectral density is evaluated from $A({\textbf{k}},\,\omega )$ at fixed frequency along nodal direction for different values of  ${k}$.

In Fig. 12 a the EDC spectral densities  at \textit{$\alpha$}=0 and  \textit{g}=0 for Fermi momentum  ${k=k_F}$  and different electron concentration (a)  and near the ${k_F}$  at \textit{n}=0.97 (b)  are shown. The EDC curves have asymmetrical character that is  in accordance with experimental results \cite{9,10}. From this figure it easy to see that the maxima  of $A({\textbf{k}},\,\omega )$ correspond to resonance frequencies in   Fig.10 a. At the same time, the maximum at ${k=k_F}$  is on the Fermi level. Notice that degree of  excitation coherence  near the Fermi level  is abruptly increased for a peculiar area of the electron spectrum where there is  a frequency change from  electron (${\raise0.7ex\hbox{$ \partial \Omega _{{\it k}}  $}\!\mathord{\left/ {\vphantom {\partial \Omega _{{\it k}}  \partial k}} \right. \kern-\nulldelimiterspace}\!\lower0.7ex\hbox{$ \partial k $}} >0$) to hole (${\raise0.7ex\hbox{$ \partial \Omega _{{\it k}}  $}\!\mathord{\left/ {\vphantom {\partial \Omega _{{\it k}}  \partial k}} \right. \kern-\nulldelimiterspace}\!\lower0.7ex\hbox{$ \partial k $}} <0$) character. Indeed, near the Fermi level the  curve 1 in Fig.12 a  is very washed out unlike the curve 3  that  corresponds the curve 3 in Fig.10 a with  a  kink   at $\Omega _{{\it k}} \sim 0$. On the other hand, the curve 2  in Fig.12 a reflects an  incoherent character of the electron excitations that is in agreement with a nearly point Fermi surface in Fig. 11 a.
\begin{figure}[h]
\begin{center}
\includegraphics[width=\columnwidth]{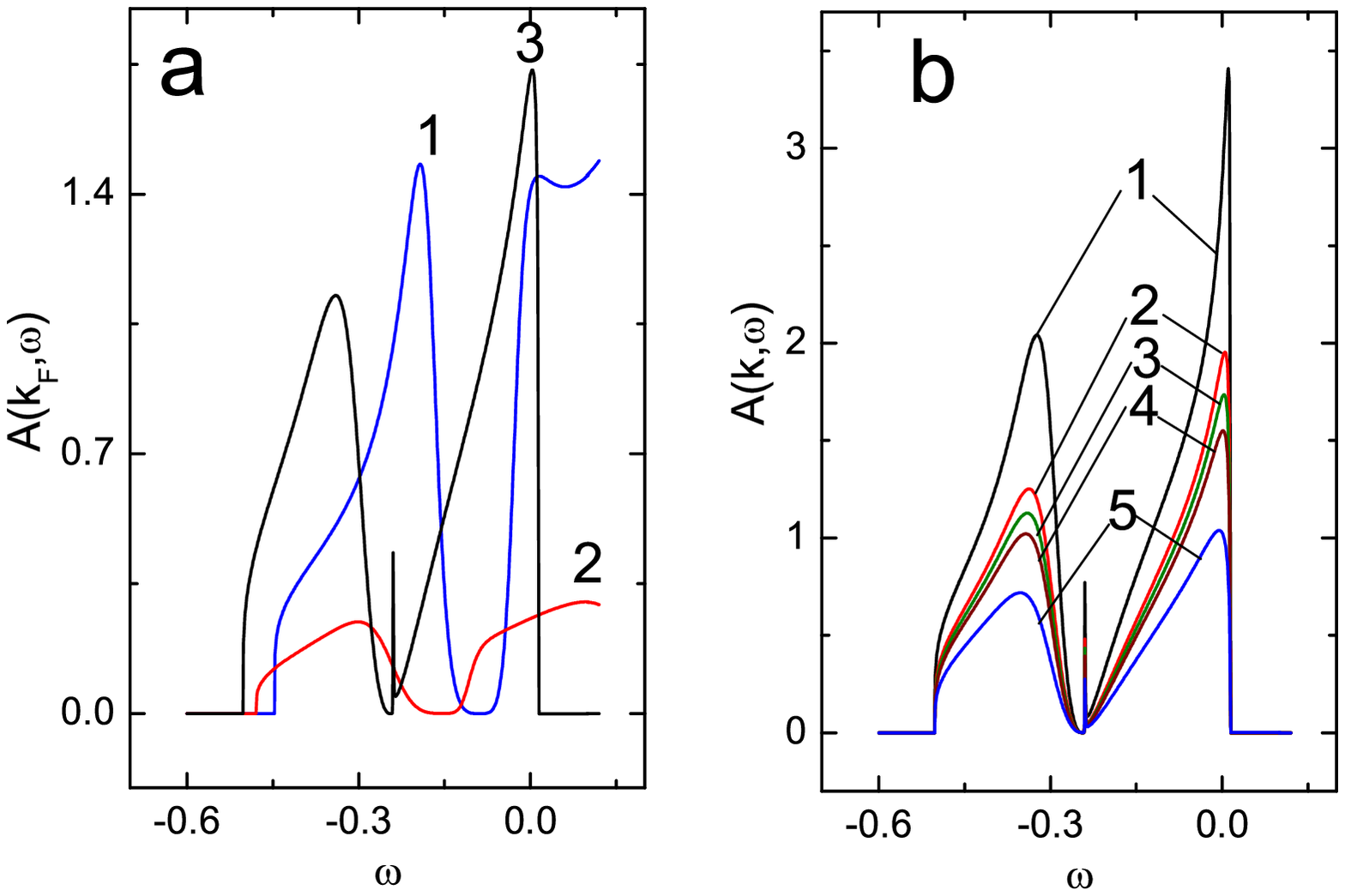}
\end{center}
\caption{EDC spectral density $A({\it k},\, \, \omega )$ from Eq. \eqref{53} as a function of the frequency at  $a=3.814 \AA$, \textit{$\alpha$}=0 and \textit{g}=0 along nodal direction of the wave vector   \textbf{\textit{k}}  with module a)  ${k=k_F}$  at  \textit{n}=0.571,  0.7  and 0.97 (curves 1-3, respectively); b) ${{k=k_F}-0.05}$, ${k_F-0.01}$, ${k_F}$, ${k_F+0.01}$ and  ${k_F+0.05}$, where  ${k_F= 0.717}$ at  \textit{n}=0.97 (curves 1-5, respectively).}
\label{fig.12}
\end{figure}

In Fig.13 the MDC spectral density $A({\it k},\, \, \omega )$ versus $\left|{\it k}\right|$ along nodal direction  at \textit{n} =0.97 for different frequencies (a) and  frequency dependence maxima of  $A({\it k},\, \, \omega )$ (b) are presented. The curve $A({\it k},\, \, \omega )$ has a Lorentz line shape that is in agreement with ARPES data. Also, we observe a  nontrivial character of the   frequency dependence maxima of  $A({\it k},\, \, \omega )$.The maximum of $A({\it k},\, \, \omega )$ appears at  ${k_{max}=0.647}$ on the Fermi level when Fermi momentum is equal 0.717. Thus, the maxima of MDC spectral density do not correspond to resonance frequencies of  the electron excitation spectrum. 
\begin{figure}[h]
\begin{center}
\includegraphics[width=\columnwidth]{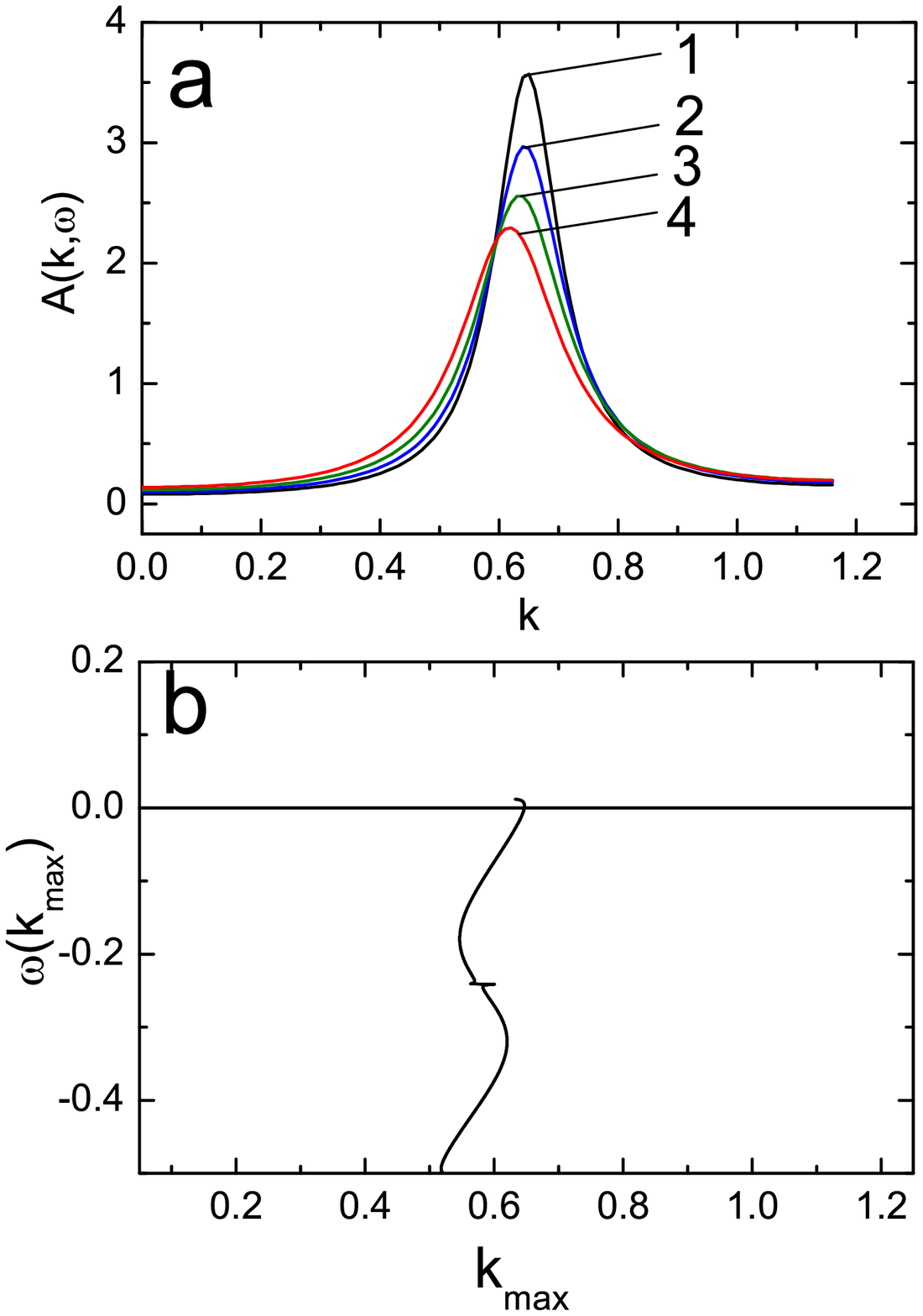}
\end{center}
\caption{MDC spectral density $A({\it k},\, \, \omega )$ from  Eq.\eqref{53} as function of a wave vector module at {$a=3.814 \AA$},  \textit{n}=0.97,  \textit{$\alpha$}=0 and \textit{g}=0  along nodal direction of  \textbf{\textit{k}}: a)   \textit{$\omega $}= 0, -0.01, -0.025 and -0.05 (curves 1-4, respectively);  b)  frequency dependence of  $A({\it k},\, \, \omega )$ maxima.}
\label{fig.13}
\end{figure}

In Fig.14 the similar dependencies of $A({\it k},\, \, \omega )$ are shown with account for an influence the next-nearest neighbors. It was considered the electron concentration \textit{n}=0.46 only when a correlation gap on the Fermi level is not developed. All conclusions to be formulated earlier  in relation to Figs.12 and 13 here remain valid. Strong coherent modes appear near  ${k\sim0.6}$. It is a result of the abrupt change of  derivation ${\raise0.7ex\hbox{$ \partial \Omega _{{\it k}}  $}\!\mathord{\left/ {\vphantom {\partial \Omega _{{\it k}}  \partial k}} \right. \kern-\nulldelimiterspace}\!\lower0.7ex\hbox{$ \partial k $}} $ (see  Fig.10.b)  for $\Omega _{{\it k}} =-\tilde{\mu }$. The edges of area of  abrupt changing $A({\it k},\, \, \omega )$ are determined by Eq.\eqref{56}  at ${\rm  }s = s_L$ and  $s_R$. It is noted that van Hove anomaly (see Fig.14 b at ${\omega \sim -0.1)}$  is determined by expression $\omega =\Omega _{{\it k0}} =-\tilde{\mu }+0.5\alpha (1-0.5n)$. It reflects a low-dimensional character of the electron behavior.
\begin{figure}[h]
\begin{center}
\includegraphics[width=\columnwidth]{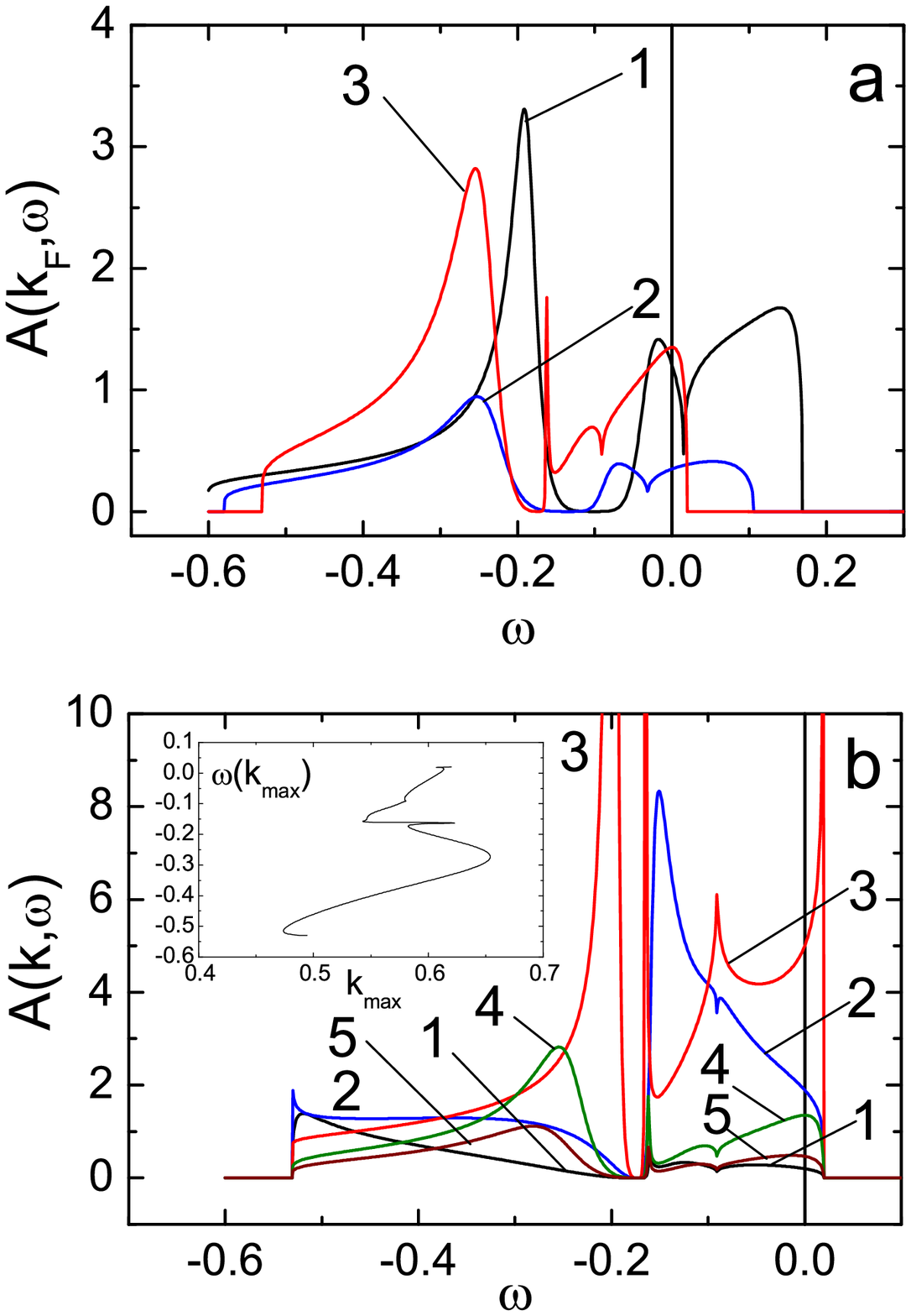}
\end{center}
\caption{EDC spectral density  $A({\it k},\, \, \omega )$ from Eq. \eqref{53} as a function of frequency  along nodal direction of vector   \textbf{\textit{k}  } at  {$a=3.814 \AA$},  \textit{g}=0,   \textit{$\alpha$}=0.3   and   a) ${k=k_F}$,  \textit{n}= 0.46, 0.63 and 0.9 (curves 1-3 for  ${k_F}$=0.734, 0.933 è 0.698, respectively);  b)  \textit{n}= 0.9  and  \textit{k}=0.4, 0.55, 0.6, 0.698 and 0.8 (curves 1-5, respectively). In inset to Fig. 13 b) frequency dependence of MDC $A({\it k},\, \, \omega )$ maxima  at  \textit{n}= 0.9.}
\label{fig.14}
\end{figure}

\subsection{\label{sec:level2} Influence of the electron-phonon interaction on electron dynamics  in cuprates}

Now we have to investigate the dynamics of the d-electrons subsystem with electron-phonon binding. First of all it is necessary to evaluate numerically the chemical potential. It was noted earlier in subsection A that diagrams 1-3 in Fig.8 give an equation for chemical potential. The solution of this equation was presented in work \cite{15}. That's why we write this equation for temperature \textit{T}=0:
\begin{eqnarray}
\label{57}
\tilde{\mu }=\frac{(2-n)}{8} M\left(1,1+\frac{\tilde{\mu }}{\omega _{0} } ,-\lambda ^{2} \right) \kern 30pt \nonumber \\
\times I^{-1} \left(1-\frac{1}{2} (1-n)\left(2-n\right),\, {\kern 1pt} \alpha \right)
\end{eqnarray}
that differs from similar Eq.\eqref{50} in that it has factor $M\left(1,1+\frac{\tilde{\mu }}{\omega _{0} } ,-\lambda ^{2} \right)$ to be equal unity in the limit case \textit{g}=0. 

In Fig.15 the chemical potential $\tilde{\mu }$ versus constant  \textit{g}  at  \textit{n }=0.9 and  \textit{$\alpha$ }=0, 0.1 and 0.3  and $\tilde{\mu }$ versus \textit{n } at  \textit{$\alpha$ }=0  and  \textit{g }= 0.01, 0.03 and 0.06 (( curves  1-3, respectively) are presented. In Fig.15 a the curves have kinks at ${n=n_{c.}}$. For ${n<n_{c.}}$ and ${n>n_{c.}}$ the  chemical potential $\tilde{\mu }$ is determined by Eq.\eqref{57} and equation  $\tilde{\mu }=-\delta \mu _{-\sigma } $, respectively. From figure it is seen that $\tilde{\mu }$ is decreased with increasing  both  \textit{$\alpha$}  and g.
\begin{figure}[h]
\begin{center}
\includegraphics[width=\columnwidth]{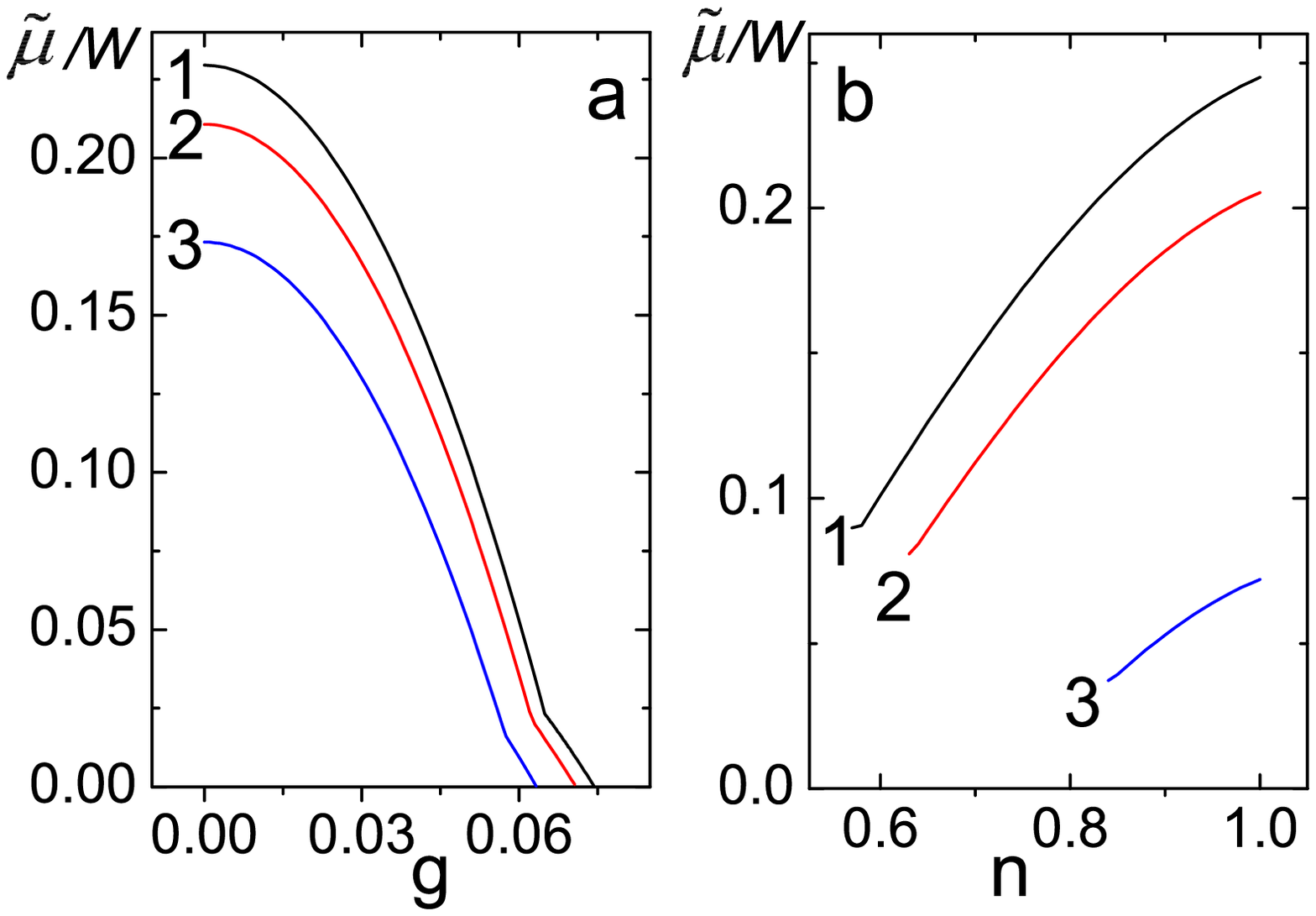}
\end{center}
\caption{a) The chemical potential versus constant of the electron-phonon interaction \textit{g} with phonon frequency $\omega _0$= 0.01875 at electron concentration   \textit{n }=0.9 and  \textit{$\alpha$ }=0, 0.1 è 0.3  (curves 1-3, respectively); b) the concentration dependencies of the chemical potential at \textit{$\alpha$ }=0  and  \textit{g }= 0.01, 0.03 è 0.06 (curves  1-3, respectively).}
\label{fig.15}
\end{figure}

Knowing the chemical potential and using Eqs.\eqref{36}-\eqref{49}, it easy  to calculate the excitation spectrum  and spectral density of the electron cuprate system. Therefore let us write the expression for  $\varepsilon _{c.} $ similar to Eq.\eqref{55} that determines the dispersion equation 
\begin{eqnarray} 
\label{58}
- \varepsilon _{c.} (\Omega _{\textbf{k}} ) =  - \frac{{4(\Omega _{\textbf{k}}  + \tilde \mu )^2 }}{{L(\Omega _{\textbf{k}} )}},  
\end{eqnarray} 
where
\begin{eqnarray}
\label{59}
L(\omega ) = (\omega  + \tilde \mu )(1 - 0.5n)M\left( {1,1 + \frac{{\omega  + \tilde \mu }}{{\omega _0 }}} \right) \nonumber \\ 
  + J_5 (\omega ) + (\omega  + \tilde \mu )^2 {\mathop{\rm Re}\nolimits} P_{46} (\omega )
\end{eqnarray}
In particular, along the nodal direction the module of $\textbf{k}$-momentum as function of  resonance frequency is written in form
\begin{eqnarray} 
\label{60}
k(\Omega _{{\it k}} )=\arccos \left(\frac{-1+\sqrt{1-2\alpha \varepsilon _{c.} (\Omega _{{\it k}} )} }{2\alpha } \right) 
\end{eqnarray} 
Apparently, the momentum  ${k_F}$  on the Fermi surface is determined from condition ${k_{F} =k(0)}$. If $-\varepsilon _{c.} (0)=-\varepsilon _{F} <0$ then we have the electron Fermi surface and hole one otherwise. A given statement may be disrupted for $\alpha\neq 0$.

In Fig.16 the dependence $-\varepsilon _{F} $ versus \textit{n} at \textit{g}=0.03 and  $\alpha$=0 is presented. It is evident from Fig.16 that  the hole Fermi surface is changed to electron one in a certain area of the electron concentrations.
\begin{figure}[h]
\begin{center}
\includegraphics[width=\columnwidth]{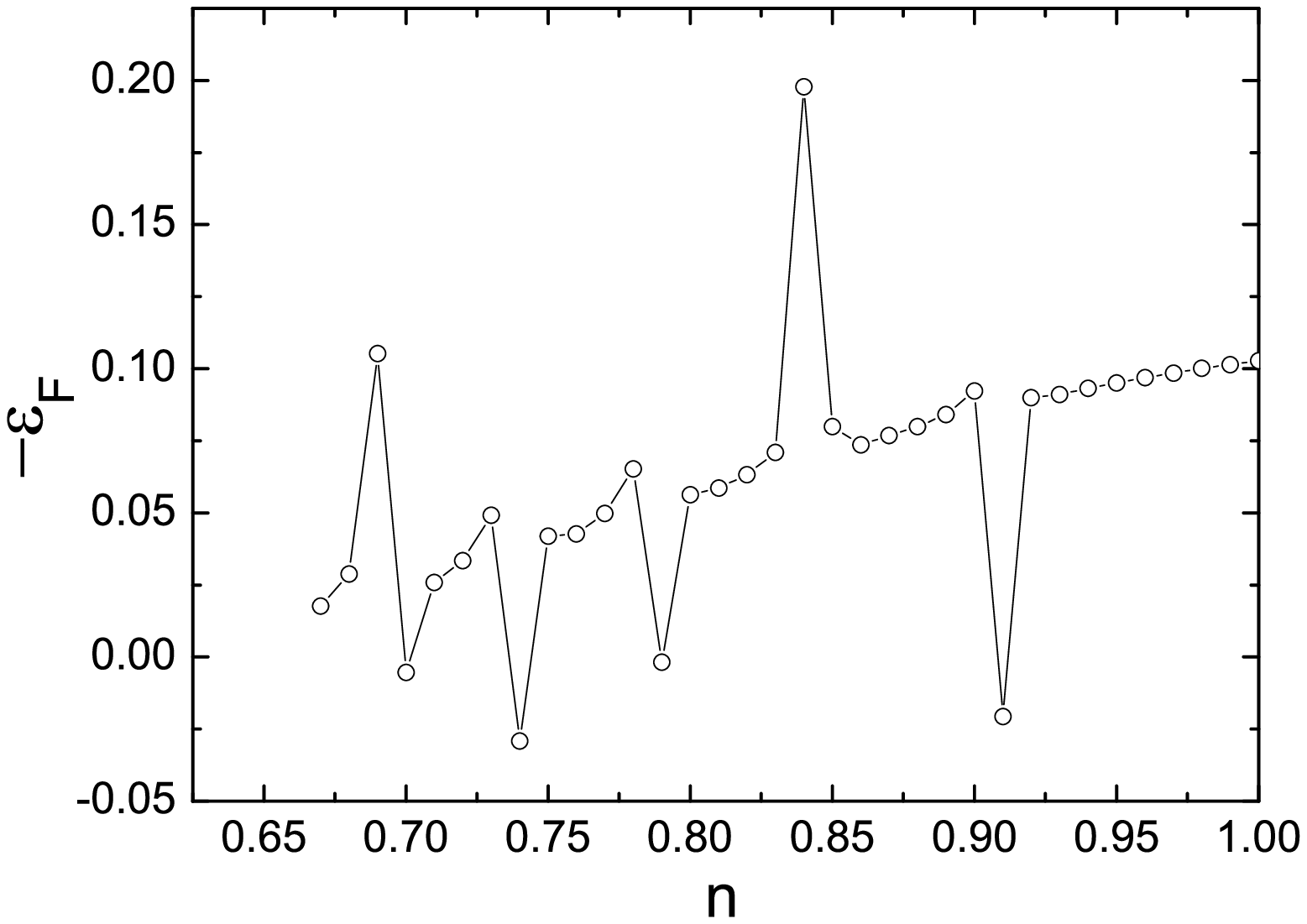}
\end{center}
\caption{The dependence of the critical parameter $-\varepsilon _{F} $ on electron concentration at  g=0.03 and  $\alpha$=0.}
\label{fig.16}
\end{figure}

In Fig.17 the electron excitation frequency-momentum dispersion curves   in a wide area of a spectrum (a) and near the Fermi level (b) along the nodal direction at \textit{n}=0.9, $\alpha$=0 for different values of \textit{g} are depicted. One can see from Fig.17 that the radical rebuilding of the excitation spectrum occurs  with increasing the electron-phonon binding energy. It is a result of the appearance of  polaron bands which are grouped near the frequencies to be multiple the phonon frequency $\omega_0$. It should be noted  that there is a weak dependence of resonance frequency on  \textit{k} at g=0.06 (see Fig.17 a, curve 4) that is in agreement with ARPES experimental data \cite{11}.
\begin{figure}[h]
\begin{center}
\includegraphics[width=\columnwidth]{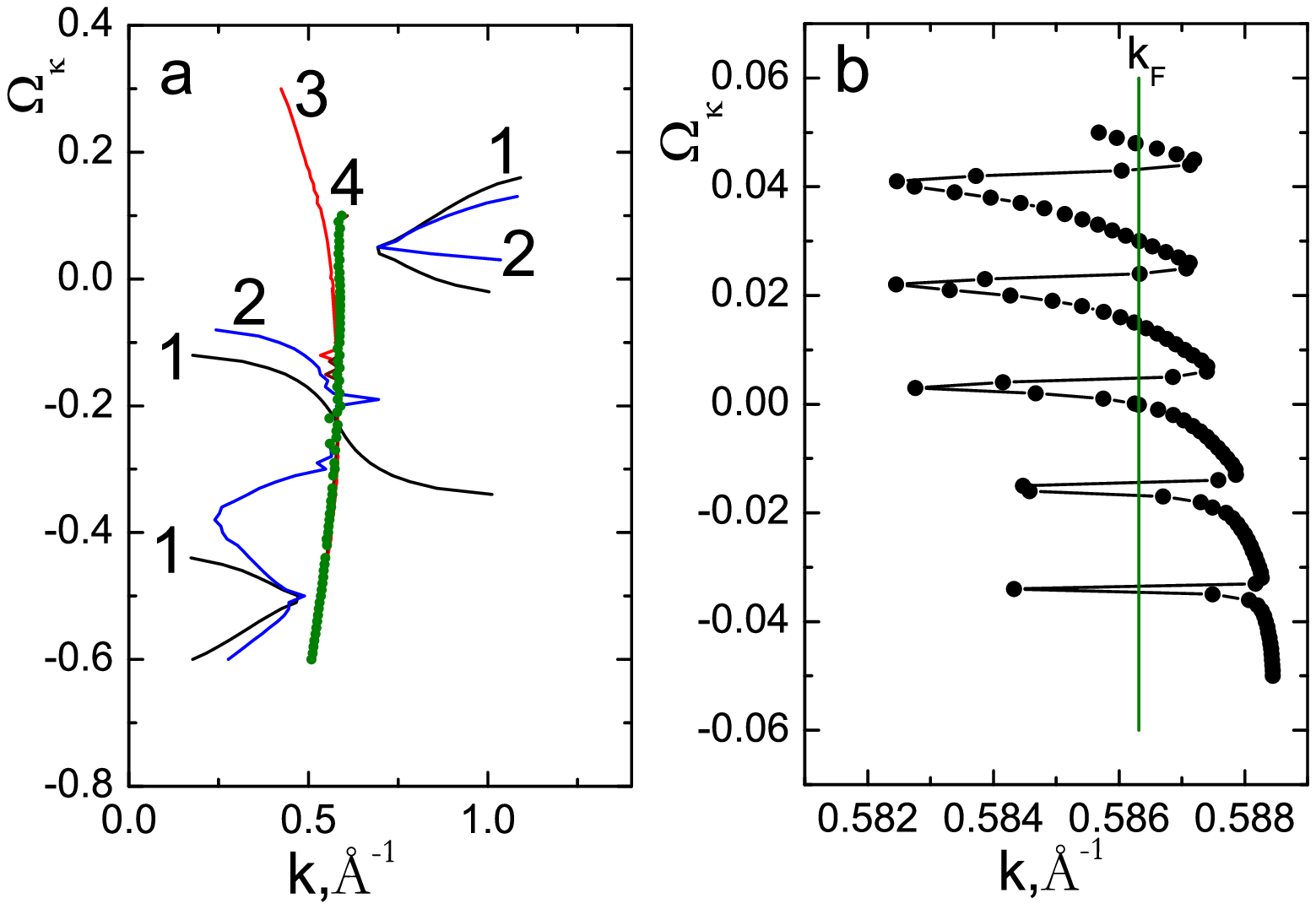}
\end{center}
\caption{a) The excitation frequency-momentum dispersion along the nodal direction  at \textit{n}=0.9, $\alpha$=0  for values of  \textit{g} =0, 0.01, 0.03 and 0.06 (curves 1-4, respectively). The points in Fig. correspond to $\alpha$=0.1 and   \textit{g}=0.06;  b)  a  fine structure of excitation spectrum  a)  at  \textit{g} =0.06. The vertical line corresponds to wave vector  ${k=k_F}$  on the Fermi surface. }
\label{fig.17}
\end{figure}

The  fine spectrum structure near the Fermi level shows an enough  drastic frequency change with a small change of \textit{k}. In particular, below the Fermi level the first kink is observed at $\Omega _{{\it k}} \sim -\omega _{0} $, i.e. at a frequency of  the typical optic phonon mode. This is in agreement with experimental data  $\Omega _{{\it k}} $${\sim}$50-70 mEv  \cite{10,24} for bandwidth ${W\sim}4 eV$. A somewhat different kink direction  from right to left  may be changed by sign replacement of hopping integral, i.e.  \textit{t} on    -\textit{t .}  Also, when the kink is observed the value of momentum ${k_{th.}\sim}$0.586 {$\AA^{-1}$} . It is in accordance with experiment ${k_{exp.}\sim}$0.41 $\AA^{-1}$ \cite{1, 9, 10}  if to put $k_{th.}$/$\sqrt{2}$. Seemingly, the maximal value of the hopping integral \textit{t} is realized along diagonal of the square lattice. Although, the other factors determining the kink position are not excluded. 

A high degree of  coherence of  excitations  near the Fermi level is the distinctive feature of  obtained spectrum with polaron bands. Indeed, for a  square lattice the excitation spectrum is very diverse at  \textit{g}=0. But the strong low-dimensional fluctuations are  responsible for a low degree of coherence of excitations excluding some edge points determined by  Im$G(s,\alpha )$. To verify this we write the expression for spectral density generally
\[
\begin{array}{l}
 A(k,\,\,\omega ) = \frac{1}{\pi } \\ 
  \times \frac{{(\omega  + \tilde \mu )^4 \beta {\mathop{\rm Im}\nolimits} P_{46} (\omega )}}{{\left[ {(\omega  + \tilde \mu )^2  - t(k)L(\omega )} \right]^2  + (\omega  + \tilde \mu )^2 t^2 (k)\left[ {(\omega  + \tilde \mu )^2 \beta {\mathop{\rm Im}\nolimits} P_{46} (\omega )} \right]^2 }} \\ 
 \end{array}
\]
where $P_{46} (\omega )$ and $L(\omega )$ are determined in subsection A  and by Eq.\eqref{59}, respectively.

In Fig.18 the EDC dependencies of  spectral density at \textit{n}=0.9, \textit{$\alpha$}=0 and  \textit{g}=0.06 ($\tilde{\mu }$ =0.053) are presented  for different momentums along the nodal direction  near the Fermi wave vector ${k_F}$=0.586 $\AA^{-1}$. From Fig.18 it is seen that the sharp peaks are observed on the curve $A({\it k},\, \, \omega )$ for Fermi momentum ${k=k_F}$ (see curve 2). Also, the satellite peaks with  smaller amplitudes are seen for other energies of excitations. These peaks correspond to points of intersection by straight line ${k=k_F}$  of polaron bands which are depicted in Fig.17 a. 
\begin{figure}[h]
\begin{center}
\includegraphics[width=\columnwidth]{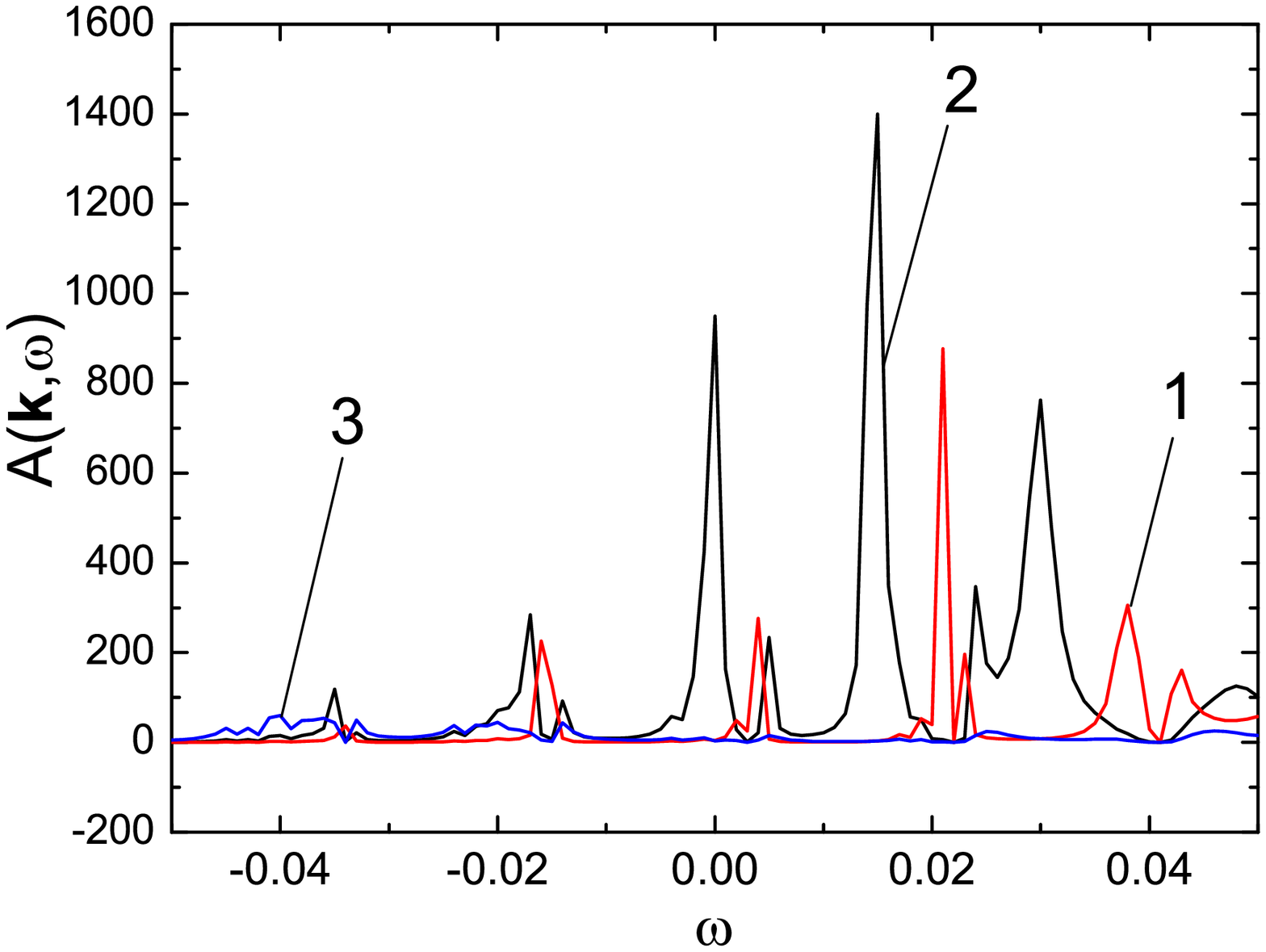}
\end{center}
\caption{The frequency dependencies of the spectral density at \textit{n}=0.9, \textit{$\alpha$}=0, \textit{g}=0.06 and         ${k=k_F}$-0.003,  ${k=k_F}$ and ${k=k_F}$ +0.003 (curves 1-3, respectively) for momentums along the nodal direction where ${k=k_F}$=0.586 $\AA^{-1}$  and parameter square lattice a${=3.814 \AA}$.}
\label{fig.18}
\end{figure}

It should  be noted  that a small change of  ${k}$ from ${k_F}$  gives a sharp decreasing of  the spectral density (see curves 1 and 3 in Fig.18). The coherence of excitations is reduced as the electron-phonon  binding  is weakened. Also, a correlation gap arises on the Fermi level for sufficiently small values of g (see Fig.17 a).

Thus, for cuprates in the strong-coupling polaron limit the hole or electron Fermi surface is realized, depending on  electron concentration and value of g. A spectral density reflects the high degree of excitation  coherence  near the Fermi level. The availability of  the narrow polaron bands, which are grouped near the frequencies to be multiple the phonon frequency $\omega _0$, results in a sharp resonance frequency changing along the nodal direction of  the momentum. It explains the origin of kink to be observed at energy $\omega\sim -\omega_0$ from ARPES data for bismuth cuprates.

\section{\label{sec:level1} Conclusions}     

For strongly correlated electron subsystem with Holstein's polarons within the framework of diagrammatic method of  perturbation theory  a generalization  for  approximation Hubbard-I  was made. Strong electron correlations narrow  a valency band  in a doped Mott insulator that results in radically difference one from metal. The Lang-Firsov unitary transform  was made to separate fermionic and bosonic subsystems. It allows to evaluate all polaron bands each of which are formed near the Einstein mode. 

In the first nonvanishing approximation of  time-dependent perturbation theory with respect to the inverse effective radius of interaction we account for influence of inelastic electron-electron scattering on chemical potential  and  time-ordered  Green's function. Taking into account a low-dimensional character of system it was obtained that all peculiarities of excitation spectrum and damping are determined by lattice Green' function for which there is a closed analytic expression. An influence of electron concentration and  next-nearest-neighbor hopping integral on chemical potential, spectrum structure  and spectral density has been determined. In particular, with well-defined electron concentrations the correlation gap in spectrum arises on the Fermi level. In the absence of this gap the Fermi surface can have the hole as well as electron character depending on electron concentration  and influence of  the  next-nearest-neighbor hopping integral. In spite of the complicated picture of excitation spectrum the spectral density shows a relatively low degree of coherence. It is connected with a low space dimension of  the system where the role of  quantum fluctuations are important.

In addition  to the previously considered approach the inclusion of  the electron-phonon interaction allowed to reveal the main peculiarities of  spectrum. They  are  in a  good correspondence to experimental ARPES data. Indeed, in a wide area of  frequencies a weak momentum dependence  of spectrum along the nodal direction was found that is in agreement with experiment [11]. The spectral density turned out to be typical for  strongly coherent excitations near the Fermi level in case of the strong  electron-phonon binding. It  simplifies  the experimental measure the Fermi surface. Thus, the electron-phonon interaction partially suppresses low-dimensional quantum fluctuations and strengthen the coherence of the excitations. From the above analysis of a fine structure of polaron spectrum one can  conclude that the observed kink  \cite{9,10}  appears at frequency of characteristic optical phonon mode in the vicinity  of  which the polaron band is formed. 

\begin{acknowledgments}
It is a pleasure to acknowledge a number of stimulating discussions with E.M. Rudenko and M.A.Belogolovsii.
\end{acknowledgments}

\end{document}